\documentclass{tcibook}
\usepackage{fancyhea}
\usepackage{work}
\usepackage{bm}       
\usepackage{graphicx}
\usepackage{hyperref}      
\usepackage[none]{hyphenat}

\newcommand{\nc}{\newcommand}  

\def\ie{{\it i.e.}}
\def\eg{{\it e.g.}}

\def\beq{\begin{equation}}
\def\eeq#1{\label{#1}\end{equation}}
\def\eeqn{\end{equation}}

\newenvironment{Eqnarray}%
   {\arraycolsep 0.14em\begin{eqnarray}}{\end{eqnarray}}
\def\beqa{\begin{Eqnarray}}
\def\eeqa#1{\label{#1}\end{Eqnarray}}
\def\eeqan{\end{Eqnarray}}

\nc{\ra}{\rightarrow}  
\nc{\slsh}{\slash\hspace*{-0.22cm}}
\def\Re{{\cal R \mskip-4mu \lower.1ex \hbox{\it e}\,}}
\def\Im{{\cal I \mskip-5mu \lower.1ex \hbox{\it m}\,}}

\nc{\vev}[1]{ \left\langle {#1} \right\rangle }
\nc{\bra}[1]{ \langle {#1} | }
\nc{\ket}[1]{ | {#1} \rangle }
\nc{\fb}{\,{\rm fb}^{-1}}
\nc{\ev}{{\rm eV}}
\nc{\kev}{{\rm keV}}
\nc{\Mev}{{\rm MeV}}
\nc{\gev}{{\rm GeV}}
\nc{\tev}{{\rm TeV}}
\nc{\mev}{{\rm MeV}}

\def\del{\partial}
\def\Dslash{\not{\hbox{\kern-4pt $D$}}}
\def\dslash{\not{\hbox{\kern-2pt $\del$}}}
\def\pslash{\not{\hbox{\kern-2pt $p$}}}
\def\ETmiss{ \not{\hbox{\kern-4pt $E$}}_T }

\def\msb{{\bar{\ssstyle M \kern -1pt S}}}

\usepackage{chngcntr}
\counterwithout{figure}{chapter}

\usepackage{amsfonts,amsmath,amssymb,amsthm}
\usepackage{enumerate}
 
\IfFileExists{srcltx.sty}{\usepackage[active]{srcltx}} 

\begin{document}

\def\bibname{References}

\bibliographystyle{utphys}  

\raggedbottom

\pagenumbering{roman}

\parindent=0pt
\parskip=8pt
\setlength{\evensidemargin}{0pt}
\setlength{\oddsidemargin}{0pt}
\setlength{\marginparsep}{0.0in}
\setlength{\marginparwidth}{0.0in}
\marginparpush=0pt


\pagenumbering{arabic}

\renewcommand{\chapname}{chap:intro_}
\renewcommand{\chapterdir}{.}
\renewcommand{\arraystretch}{1.25}
\addtolength{\arraycolsep}{-3pt}


\renewcommand*\thesection{\arabic{section}}

\chapter*{Snowmass-2013 Cosmic Frontier 3 (CF3) Working Group Summary: \\ Non-WIMP dark matter}

{\large \bf Conveners: Alexander Kusenko, Leslie J. Rosenberg}

\section*{Acknowledgments}

\begin{sloppypar}  
\large{ 
This report was based on a broad input from the scientific community.  We gratefully acknowledge contributions in the form of 
written text, figures, oral presentations, and comments, which we received from 
{\bf 
K.~Abazajian,  
E.~Adelberger, 
T.~Asaka, 
K.~J.~Bae, 
H.~Baer, 
K.~Baker, 
D.~Boyanovsky, 
E.~Bulbul,  
G.~Carosi, 
C.~Charlett, 
C.~Cheung,  
A.~Chou, 
A.~Cieplak, 
D.~Cline, 
H.~de Vega, 
K.~Dienes, 
S.~Dodelson, 
R.~Essig, 
J.~L.~Feng, 
A.~Friedland, 
G.~Fuller, 
M.~Giannotti, 
P.~Graham, 
K.~Griest, 
P.~Gondolo,  
D.~Kaplan, 
M.~Kaplinghat,  
P.~Karn, 
J.~Kim, 
A.~Kratsov, 
J.~Kumar, 
S.~Lamoreaux,  
A.~Linder, 
M.~Loewenstein, 
J.~Mardon, 
J.~Merten, 
T.~Moroi, 
L.~Moustakas,  
G.~Mueller,  
A.~Nelson, 
Y.~Nomura, 
L.~Pearce, 
R.~Peccei, 
K.~Petraki, 
M.~Pivovaroff,  
C.~Pfrommer,  
G.~Raffelt, 
J.~Redondo, 
A.~Ringwald, 
G.~Rybka, 
J.~Ruz,  
T.~Quinn, 
C.~Reichardt,  
O.~Ruchaiskiy, 
P.~Sallucci,  
M.~Sanchez-Conde, 
I.~Shoemaker, 
R.~Shrock, 
J.~Siegal-Gaskins, 
K.~Sigurdson, 
P.~Sikivie,  
F.~D.~Steffen, 
L.~Strigari, 
T.~Tait, 
D.~Tanner, 
B.~Thomas, 
J.~Tiffenberg, 
K.~van Bibber,
J.~Vogel,
R.~Volkas,
M.~Walker, 
W.~Wester, 
F.~Wilczek,    
K.~Zioutas,
A.~Zhitnitky, 
K.~Zurek.
}} 
\end{sloppypar}

\tableofcontents

\section{Introduction}

In solving the mystery of dark matter, one sensible way to proceed is to assume the properties of the dark-matter particle candidate and predict its non-gravitational interactions 
by means of which the particle can be identified.  It is common to predict the properties of dark matter based on either compelling theoretical arguments or 
experimental and astrophysical hints.  The relative significance of different arguments and hints cannot be evaluated objectively, which therefore
makes it difficult to  rank dark matter candidates in importance.   While a combination of certain theoretical arguments, advanced direct-search experimental techniques, 
and connection with collider experiments makes Weakly Interacting Massive Particles (WIMPs) a very appealing candidate, it is by no means the only
or most appealing possibility.
The {\em non-WIMP dark matter} subgroup ``CF3''  of the Cosmic Frontier was charged with a study of a broad range of dark matter candidates that do not 
fall into the category of WIMPs.

The search for physics beyond the standard model has been shaped by aesthetic considerations.  {\em Occam's razor} reasoning and naturalness arguments are 
often invoked.  Some of the proposed candidates for dark matter stand out because they are motivated by strong independent reasons, which makes them 
appear more plausible than {\em ad hoc} solutions to the dark matter problem. The axion is an example of a
compelling dark-matter candidate motivated by strong independent arguments.

However, one must exercise some caution in applying aesthetic arguments for prioritizing candidates. Many great discoveries have revealed the properties of nature 
that were not considered ``natural'' by most researchers prior to their discoveries.  Arguably, a universe without dark energy and without dark matter would be a simpler and more 
Occam-friendly universe, but major discoveries in observational cosmology have proven some common theoretical prejudices incorrect.  In particle physics, 
the existence of three generations of fermions presents another challenge to  {\em Occam's razor}.  In fact, we find it surprising how few are the examples of successful 
aesthetic arguments leading to a discovery, in particle physics or in any other branch of science.  Naturalness will play a role in our discussion, but it will 
not be used as the ultimate litmus test for a successful dark matter candidate.

Another important consideration for planning the future research is feasibility.  An experimental program only makes sense if it is capable of achieving its scientific 
goals using available technologies.  If two dark matter candidates are equally plausible based on current knowledge, but only one of them is readily amenable to detection, 
it is well justified to direct more resources to the pursuit of the discoverable candidate.   The search for one's keys under the lamp post is perfectly justified if 
there is no reason to favor one location over the other, based on the best information available.

This report will not attempt to discuss all possible dark matter candidates and experimental techniques, which is, clearly, impossible in such a
short discussion.  We will present a limited but, hopefully, balanced overview of the possible opportunities in the search for the identity of dark matter.

\section{Theoretical motivation}

The deductive approach is limited in application to the dark matter problem, and it is probably fair to conclude that one must guess the answer before one can identify the dark matter particle(s).  For each candidate, one tries to elucidate the possible interactions and to design detection strategies suitable for such a particle~\cite{Feng:2010gw}.  A broad, comprehensive, and open-minded approach to a wide range of candidates is much more likely to succeed than a narrow pursuit of one or a few possibilities.   Furthermore, there is no reason to believe that dark matter is comprised of only one type of particle. 
It is possible that that the structure of the dark sector is as complex as that of the visible sector, or it may have an even reacher structure.  For example, dynamical dark matter models~\cite{Dienes:2011ja,Dienes:2011sa} predict a large number of species in the dark sector. 

\section{General discussion of dark matter properties}  

The number of basic requirements for dark matter are very few.  The dark-matter particles must be stable, or at least stable on time scales much longer than the age of today's universe.  These particles must be produced in the early universe (which usually compels one to hypothesize some interactions, although in some cases gravitational interactions can suffice).  Finally, they must form cosmological structures consistent with astronomical observations. The latter requirement is satisfied by collisionless (non-interacting) cold (having a negligible primordial velocity dispersion) dark matter (CDM), although significant discrepancies exist between structures predicted in N-body simulations and observations.  These discrepancies may be shortcomings of the simulations, or they may indicate that dark matter is not entirely collisionless CDM.  Furthermore, even if future improved simulations come in perfect agreement with observations, they will still leave room for a component of dark matter that is not 
collisionless.  

Dark matter self-interaction can have a profound effect on the shape of the density profile because additional (besides gravity) interactions of dark matter particles can facilitate the transfer of momentum and angular momentum through the halo.  The relevant quantity is not the cross section or the  mass, but the ratio of self-interaction cross section to the mass of the particle, $\sigma/m$.  The larger the mass, the lower is the number density, and, therefore, a larger cross section is needed to generate some
non-negligible number of self-interactions.  
Therefore, the heavier the particle, the larger is the allowed cross sections of self-interactions. Some of the candidates discussed below have very
large mass and very large cross sections, indeed.

Dark matter interactions with ordinary matter must not be significant enough for dark matter to collapse into galactic discs, but this requirements still allows
heavy enough particles to have strong interactions (QCD) or electromagnetic interactions, for example. 
Heavy dark matter particles can be present in low enough density to evade detection even if they have electric charge.   
This applies only to the heaviest of all candidates, but the fact remains that such interactions are not, in general, ruled out.

In summary, dark matter particles with a broad range of possible interactions are still allowed by astrophysics and by experiment.

\subsection{Astrophysical observations and insights}  
\label{sec:astro}
While collisionless CDM is a good reference-point type of dark matter which reproduces large-scale structure, some persistent small inconsistencies between the predictions of N-body simulations and the observations may provide unexpected new insights into the properties of dark matter particles.   Such deviations can be indicative of self-interactions or primordial velocity distribution different from those of CDM.  At present, none of these 
discrepancies provide a smoking-gun evidence for any particular dark-matter candidate.  However, there is a tantalizing possibility that astronomical
observations  and astrophysical insights can point to specific properties of dark-matter particles.   A separate section of the Cosmic Frontier report is devoted to a detail discussion of astrophysical probes that can reveal the microscopic properties of dark-matter particles.

\section{The (incomplete) landscape of candidates}

\begin{figure}[!htb]
\begin{center}
  \includegraphics[height=0.4\textheight]{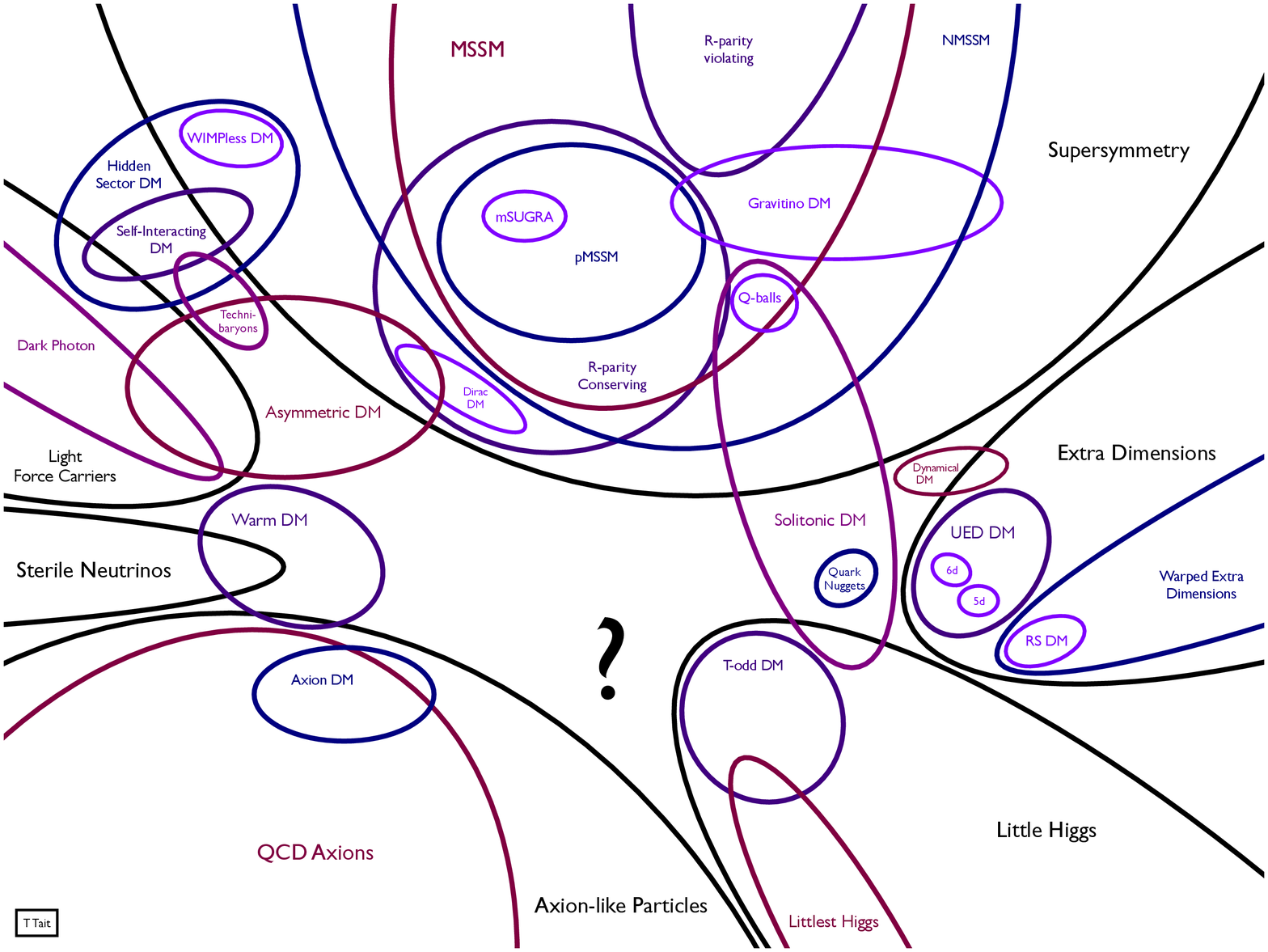} \\
  \includegraphics[height=0.4\textheight]{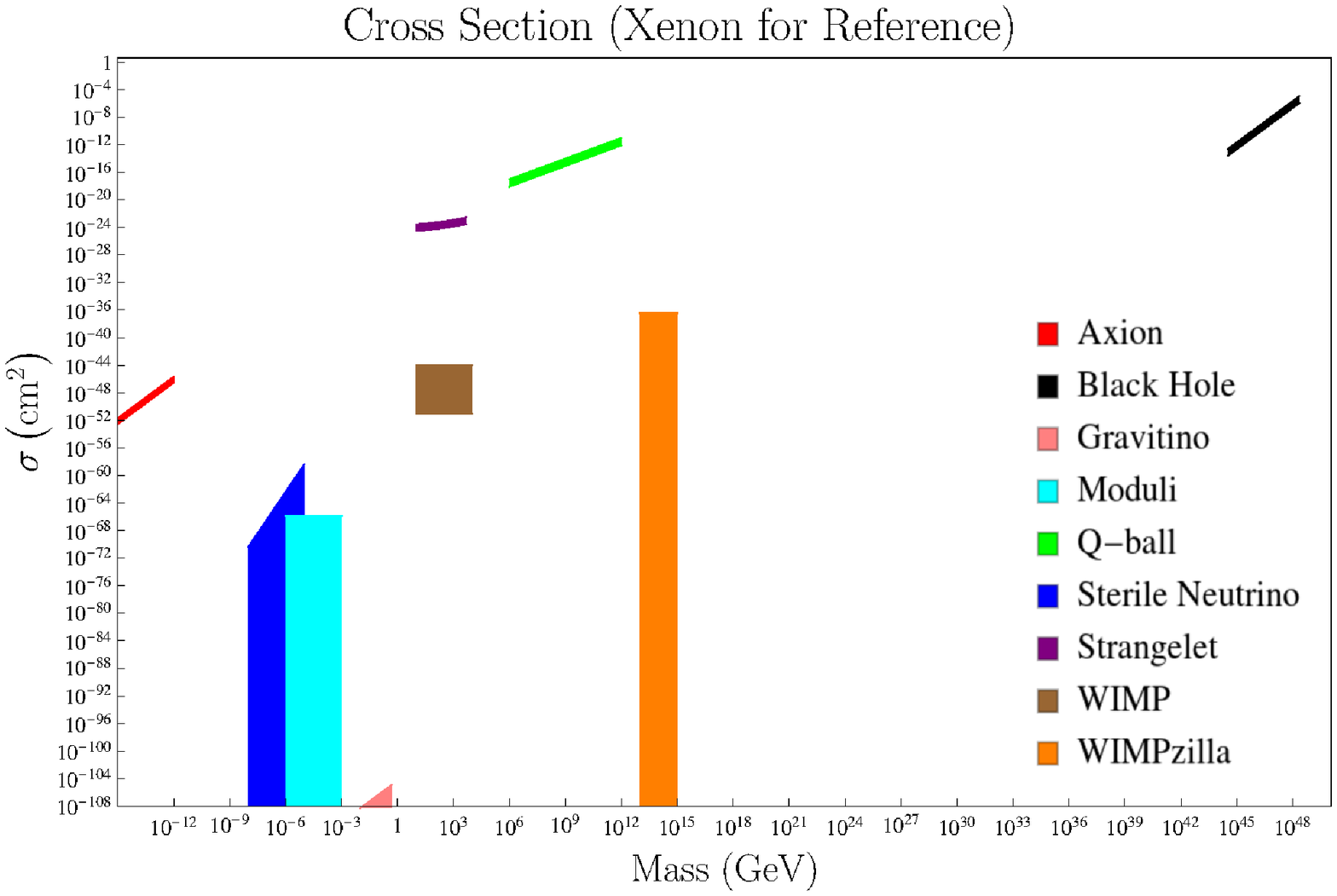}
\caption{Graphical representation of the (incomplete) landscape of candidates.  Above, the landscape of dark matter candidates due to T.~Tait.  Below, the range of dark matter candidates' masses and interaction cross sections with a nucleus of Xe (for illustrative purposes) compiled by L.~Pearce. Dark matter candidates have an enormous
range of masses and interaction cross sections.}
\end{center}
\end{figure}

The following sections of this report discuss some of dark matter candidates in more detail.

\subsection{Asymmetric dark matter} 
\label{sec:ADM}

Models of Asymmetric Dark Matter (ADM) are based on the idea that the dark matter, like the baryons, may carry a matter-anti-matter asymmetry. 
This appealing paradigm is reviewed in detail in Refs.~\cite{Petraki:2013wwa,Zurek:2013wia}.

The idea that there may be a relation between the dark matter and baryon asymmetries was proposed a long time ago~\cite{Nussinov:1985xr}.  Many of the earliest models (see, for example, Refs.~\cite{Barr:1990ca,Barr:1991qn,Kaplan:1991gb,Gudnason:2006ug,Gudnason:2006yj}) made use of electroweak sphalerons to distribute baryon and lepton number between the visible and dark sectors; such models have subsequently become highly constrained by both LEP and the LHC.  A recently proposed approach~\cite{Kaplan:2009ag} provided a flexible and robust framework to relate the baryon and dark matter via higher dimension operators, which easily evades the constraints from electroweak scale dynamics.  These higher dimension operators naturally decouple at low temperatures, separately freezing in the dark matter and baryon asymmetries.  

As discussed below, ADM can naturally accommodate relatively low masses of dark matter particles, around a few GeV.  ADM models received a lot of attention in light of several direct detection anomalies suggesting a low mass WIMPs in the in sub-10 GeV range.  Mirror dark matter models \cite{Petraki:2013wwa} and hidden sector dark matter models (an explicit construction can be found in \cite{Cohen:2010kn}) have been motivated in part by these signals.  In addition, the need to annihilate the symmetric component of the dark matter motivates the presence of light dark forces, which may be searched for in low energy $e^+ e^-$ experiments~\cite{Izaguirre:2013uxa}.  The presence of light dark forces implies dark matter self-scattering, which has potentially important implications for structure formation.

\subsubsection{Motivation and general features}

The similarity of the visible matter (VM) and the DM relic densities suggests the possibility of a common origin. If instead they originated via unrelated mechanisms, their values must have been determined by different fundamental and cosmological parameters, and would be expected generically to differ by many orders of magnitude. However, precision measurements of the cosmic microwave background reveal that~\cite{Hinshaw:2012aka,Ade:2013zuv}
\begin{equation}
\Omega_{\rm DM} \simeq 5 \, \Omega_{\rm VM} \, ,
\end{equation}
where $\Omega_i$ denotes the present-day energy-density fraction of the $i$-th component of the universe. The primary motivation of the Asymmetric DM (ADM) scenario is to offer a dynamical explanation for this cosmic coincidence of visible and dark matter. 
In fact, there are more hints pointing towards a connection between the physics of VM and DM. 
One interpretation of claimed signals from DM direct-detection experiments point to a DM mass scale rather similar to the nucleon mass, in the few GeV range~\cite{Bernabei:2008yi,Bernabei:2010mq,Aalseth:2010vx,Aalseth:2011wp,Angloher:2011uu,Agnese:2013rvf}. 
The observed clustering patterns of DM can be explained better by DM with self-interaction cross-section within an order of magnitude from the neutron self-scattering
cross-section, rather than by collisionless cold DM.
These features arise naturally in ADM models and encourage this line of investigation. 
Moreover, within the phase space of DM models, the ADM paradigm encompasses the entire continuum of thermal-relic DM with annihilation cross-sections larger than the canonical value for symmetric WIMP DM. This is obviously independent of whether a connection between VM and DM is established, and includes a variety of models which possess independent motivations.

To begin contemplating how the physics and the cosmological evolution of VM and DM may be related, we first consider some of the relevant properties of VM.
It has been long established that the relic abundance of VM is due to an excess of matter over antimatter in the today's universe. The evidence is two-fold: The lack of abundant particle-antiparticle annihilation signatures in astrophysical signals and the very small fraction of antimatter in cosmic rays manifest the near absence of antimatter from the observable universe. Moreover, the known properties of VM imply that the matter-antimatter annihilation processes in the early universe would have been too rampant to allow for a relic abundance of VM as large as we observe today, if equal amounts of matter and antimatter were present in the universe. 
On the other hand, in an expanding universe with asymmetric particle-antiparticle content, annihilations cannot diminish the relic density of matter below the existing asymmetry, provided of course that the fundamental interactions distinguish between particles and antiparticles and preserve their net number. Indeed, the particle-antiparticle asymmetry in VM, known as the baryon asymmetry of the universe (BAU), is maintained in the low-energy environment of today's universe due to the baryon-number symmetry of the Standard Model (SM). 
However, the origin of the BAU cannot be adequately accounted for by the SM processes and is still unknown. The dynamical generation of the BAU, known as baryogenesis, requires us to suppose interactions which satisfy the three Sakharov conditions~\cite{Sakharov:1967dj}: 
First, these interactions must violate the baryon-number symmetry, and they must have been effective in the high-energy environment of the early universe, but ceased as the universe expanded and cooled. Second, they must violate the discrete C and CP symmetries.
Third, they must have occurred out of equilibrium, such that the reverse processes could not have washed out the generated baryon number.

Relating the BAU to the DM relic density likely implies some connection between baryogenesis and the processes which established the relic abundance of DM.
The ADM hypothesis states that the DM relic abundance is also due to a particle-antiparticle asymmetry in a dark particle number, and that this asymmetry was dynamically linked to the BAU by processes that occurred in the early universe. There are three essential elements in this scenario: 
\begin{enumerate}[(a)]
\item 
There must be a global additive particle-number symmetry governing the low-energy interactions of DM, such that the dark matter-antimatter asymmetry is preserved today.
We shall call this symmetry ``dark baryon number'' and denote it by $B_{\rm D}$, in analogy to the baryon number of the SM which we shall denote by $B_{\rm V}$. DM is envisioned to be the lightest particle transforming under $B_{\rm D}$.
\item 
There exist high-energy interactions which violate a non-trivial combination of $B_{\rm D}$ and $B_{\rm V}$ while preserving the orthogonal number. These processes must have operated in the early universe, relating the $B_{\rm V}$ and $B_{\rm D}$ asymmetries, and must have become inefficient at later times, allowing for the sequestration of the two global charges.
(Note that due to the sphaleron effects of the SM which link $B_{\rm V}$ to the lepton-number symmetry of the SM, $L_{\rm V}$, we can consider $(B-L)_{\rm V}$ instead of $B_{\rm V}$ in the above.)

\item 
Dark antibaryons annihilated efficiently in the early universe via an appropriate interaction, leaving the excess dark baryons as the dominant component of DM today.
\end{enumerate}

Below we expound on these elements, which provide the framework for ADM model building, and also determine the generic features of the low-energy phenomenology of ADM models. As we shall see, the ADM scenario can accommodate a large range of possibilities for the properties of DM, with markedly different scenarios in how DM clusters and what direct and indirect signals it is expected to produce. 
For reviews and complete lists of references, see Refs.~\cite{Petraki:2013wwa,Zurek:2013wia,Davoudiasl:2012uw}.

\paragraph{Symmetry structures and symmetry breaking patterns of ADM models. }

There are two qualitatively different symmetry structures which allow for the VM and DM particle asymmetries to be related.
\begin{enumerate}[(i)]
\item
The visible and the dark baryonic asymmetries are generated simultaneously via the same interactions. 
We may consider two non-trivial and linearly independent combinations of the low-energy symmetries
\begin{equation}
\begin{alignedat}{5}
& B_{\rm con} &  \ = \ & B_{\rm V} - B_{\rm D} & \ , \\
& B_{\rm bro} &  \ = \ & B_{\rm V} + B_{\rm D} & \ .
\end{alignedat}
\end{equation} 
In this class of ADM models, $B_{\rm con}$ remains (effectively) unbroken throughout the cosmological evolution of the universe. The linearly independent combination, $B_{\rm bro}$, is broken in the early universe by high-energy interactions which also violate C and CP and occur out of equilibrium. With the Sakharov conditions satisfied for the number $B_{\rm bro}$, a net charge $\Delta B_{\rm bro}$ is generated. By virtue of conservation of $B_{\rm con}$, this amounts to equal dark and visible baryonic asymmetries, 
$\Delta B_{\rm V} = \Delta B_{\rm D} = \Delta  B_{\rm bro} /2$.
After the asymmetry generation, the $B_{\rm bro}$-violating interactions decouple, and both $B_{\rm V}$ and $B_{\rm D}$ are restored as independent symmetries of the interactions. The global charges $\Delta B_{\rm V}$ and $\Delta B_{\rm D}$ cascade down to the lightest particles transforming under the corresponding symmetries, by $B_{\rm V,  D}$ -preserving interactions.
This scenario gives rise to what has been called a  ``baryon-symmetric'' universe~\cite{Dodelson:1989cq,Dodelson:1989ii}, in the sense that the universe bears no asymmetry under a well-defined generalized baryon number, $B_{\rm con}$. It has been implemented in a number of different models which employ different asymmetry-generation mechanisms and/or different dark sector microphysics~\cite{Kuzmin:1996he,Kitano:2004sv,Kitano:2005ge,Gu:2009yy,Davoudiasl:2010am,Gu:2010ft,Heckman:2011sw,Gelmini:1986zz,Farrar:2005zd,Hall:2010jx,Bell:2011tn,Cheung:2011if,Graesser:2011vj,vonHarling:2012yn,MarchRussell:2011fi,Petraki:2011mv,Cheung:2013dca}.

\item 
An initial asymmetry is generated in $B_{\rm V}$ or $B_{\rm D}$ by high-energy processes which violate the corresponding number along with C and CP, and occur out of equilibrium. It is also possible that initial but unrelated asymmetries are generated in both  $B_{\rm V}$ and $B_{\rm D}$. At intermediate energies, processes which respect only one non-trivial linear combination of $B_{\rm V}$ and $B_{\rm D}$ get in equilibrium. The initial asymmetry/ies  is/are then redistributed via chemical equilibrium among particles of the two sectors. The decoupling of these interactions leaves a net global charge in each sector, preserved separately by the low-energy interactions. The details of the chemical equilibrium (relativistic/non-relativistic, number of species, baryonic charges) determine the exact relation between the final visible and dark baryonic asymmetries~\cite{Blinnikov:1982eh,Blinnikov:1983gh,Foot:1991bp,Foot:1991py,Foot:1995pa,Berezhiani:2000gw,Berezhiani:2003wj,Ignatiev:2003js,Foot:2003jt,Foot:2004pq,
Hooper:2004dc,Ciarcelluti:2004ik,Ciarcelluti:2004ip,Berezhiani:2005vv,Dutta:2006pt,Berezhiani:2008zza,Kaplan:2009ag,Cohen:2009fz,Cai:2009ia,Blennow:2010qp,Dutta:2010va,Buckley:2010ui,Shelton:2010ta,Haba:2010bm,Falkowski:2011xh,Haba:2011uz,Servant:2013uwa,Feng:2013wn,Foot:2013msa,Mitropoulos:2013fla}.

\end{enumerate}

\paragraph{Freeze-out in the presence of an asymmetry. }
Dynamically relating the particle-antiparticle asymmetries in the visible and the dark sectors yields a tight relation between the relic number densities of VM and DM, provided that the excess of particles over antiparticles is the only relic from each sector contributing significantly to the total energy density. In the visible sector, the $B_{\rm V}$-preserving interactions of the SM efficiently annihilate the ordinary antibaryons in the early universe, leaving only the excess of baryons present today. 
Similarly, the symmetric part of DM is efficiently annihilated away in the early universe if there exist sufficiently strong $B_{\rm D}$-preserving interactions that allow DM annihilation.

How large the DM annihilation cross-section has to be in successful ADM models is, of course, an important feature which determines their low-energy phenomenology. We may define the fractional asymmetry of DM
\begin{equation}
r \equiv n(\bar{\chi})  /  n(\chi) \ ,
\label{eq:r}
\end{equation}
where $n(\chi), \ n(\bar{\chi})$ stand for the number densities of DM particles and antiparticles respectively. Obviously $r = 0$ and $r=1$ correspond to the completely asymmetric and symmetric case respectively. The detailed Boltzmann-equation analysis shows that the late-time fractional asymmetry, $r_\infty,$ depends exponentially on the annihilation cross-section~\cite{Graesser:2011wi} 
\begin{equation}
r_\infty 
\simeq \exp \left[-2 \left( \frac{\sigma_{_0}}{ \sigma_{_\text{0,WIMP}}} \right) \left(\frac{1-r_\infty}{1+r_\infty} \right) \right] 
\quad \xrightarrow{r_\infty \ll 1} \quad
\exp \left( -2 \sigma_{_0} / \sigma_{_\text{0,WIMP}}\right )   
\  ,
\label{eq:FractionalAsymmetry}
\end{equation}
where $\sigma_{_0}$ is related to the thermally-averaged cross-section times velocity,
$\langle\sigma v \rangle = \sigma_{_0} (T/m{_{_{\rm DM}}})^n$,
with $n=0\text{ and }1$ for $s$-wave and $p$-wave annihilation respectively. The subscript ``WIMP" refers to the cross-section needed to produce the observed DM abundance via freeze-out, in the case symmetric and non-self-conjugate DM, $\langle\sigma v \rangle_{_\text{WIMP}} \simeq 6 \cdot 10^{-26} \text{cm}^3/\text{s}$.

Because of this exponentially sensitive dependence,  $\sigma_{_0} \gtrsim 1.4 \, \sigma_{_\text{0,WIMP}}$ suffices to render $r_\infty \lesssim 0.1$. Thus, annihilating efficiently the symmetric part of DM in ADM models requires an annihilation cross-section which is larger than the canonical thermal-relic annihilation cross-section, albeit only by a factor of a few~\cite{Graesser:2011wi,Iminniyaz:2011yp},
\begin{equation}
\sigma \gtrsim \text{ few } \times \sigma_{\rm _{WIMP}} \  .
\label{eq:sigma ann}
\end{equation}
Conversely, any thermal DM candidate for which Eq.~\eqref{eq:sigma ann} holds true, has to possess a particle-antiparticle asymmetry in order to account for the observed DM abundance, and the DM relic abundance will be predominantly asymmetric. (This holds even in non-standard cosmologies, albeit with $\sigma_{_\text{WIMP}}$ having a different value than in the standard cosmology.) In this sense, the ADM scenario encompasses the entire range of thermal-relic DM models with annihilation cross-sections larger than the canonical value, independently of whether a connection of the DM asymmetry with the BAU is pursued.

\paragraph{Dark interactions.} Driven by Eq.\eqref{eq:sigma ann}, it is important to consider the possibilities for DM annihilation in the ADM scenario. Efficient annihilation is obviously an essential feature of thermal-relic DM, be it symmetric or asymmetric. It is thus the feature which largely sets the expectations for experimental detection. Annihilation of DM directly into SM particles is probed in colliders, direct-detection experiments and indirect searches, with current experimental results having severely constrained the relevant parameter space~\cite{Bai:2010hh,Buckley:2011kk,Fox:2012ee,MarchRussell:2012hi}. While DM annihilation cross-sections $\sigma_{\chi \bar{\chi} \to \text{SM}} \gtrsim \sigma_{_\text{WIMP}}$ are still possible, the existing stringent bounds suggest that DM might in fact annihilate predominantly into non-SM species, which could subsequently either decay into SM degrees of freedom or constitute dark radiation.
Because there is no upper bound, but rather only a lower bound on the DM annihilation cross-section in the ADM scenario, ADM can comfortably have sizable couplings directly to light degrees of freedom. 
In this case, the couplings which provide large enough annihilation cross-section, often also provide significant DM self-interaction inside halos. Self-interaction via light mediators can result in  (velocity-dependent) self-scattering cross-sections which can play an important role in resolving the galactic structure-formation problems of collisionless CDM while preserving its successful predictions. Moreover, if the light degrees of freedom are stable, they can contribute to the relativistic energy density of the universe with observable consequences.

A simple possibility for efficient annihilation is that the dark baryons are charged under a dark Abelian gauge force, $U(1)_{_D}$, which may be broken or unbroken. Dark baryons can annihilate either directly into dark gauge bosons, provided that this is kinematically allowed, or via dark gauge bosons into other light species charged under $U(1)_{_D}$. 
If $U(1)_{_D}$ is unbroken, or if it breaks after the processes which determine the relic dark baryonic asymmetry have decoupled, gauge invariance implies that DM must consist of (at least) two species of particles, charged oppositely under $U(1)_{_D}$. 
This is because, similarly to the situation with ordinary matter, the dark baryonic asymmetry corresponds also to a net dark gauge charge carried by the dark baryons, which has to be compensated by an asymmetric population of particles oppositely charged under the gauge symmetry. Depending on the gauge coupling, the oppositely charged species can bind and form $U(1)_{_D}$-neutral atomic states. The potential phenomenological implications of atomic DM are rich, and are partly discussed in section~\ref{sec:self-interDM}. 
If $U(1)_{_D}$ is broken at a scale higher than the scale of dark baryogenesis, then it is possible that DM consists only of the dark baryons, which interact with each other via a repulsive Yukawa potential. (However, it is still possible that there are two stable species with opposite $U(1)_{_D}$, one stabilized by $B_D$ and one stabilized by a global remnant of $U(1)_D$, and that the universe has zero net charge under $U(1)_D$.)
A dark Abelian gauge force can mix kinetically with the hypercharge~\cite{Holdom:1985ag}
\begin{equation}
\delta {\cal L} = \frac{\epsilon}{2} \: F_{Y  \mu\nu} \, F_D^{\mu\nu}  \  ,
\label{eq:kinetic mixing}
\end{equation}
with a number of observable consequences, \eg~for direct detection~\cite{Foot:2003iv,Foot:2010hu,Foot:2011pi,Foot:2013msa,Foot:2012cs,Fornengo:2011sz} and dark-force experiments~\cite{Badertscher:2003rk,Jaeckel:2013ija}. If the dark photon is massless, it contributes to the relativistic energy density of the universe. If it is massive, it can decay via the kinetic mixing into SM fermions, with lifetime
\begin{equation}
\tau_{_D} \approx 10^{-13}  \text{ s} 
\left( \frac{10^{-4}}{\epsilon} \right)^2 
\left( \frac{100 \text{ MeV} }{M_D} \right) \ ,
\label{eq:dark photon lifetime}
\end{equation}
which can be sufficiently short to ensure no extra radiation contribution at late times.

Other possibilities for DM annihilation into non-SM particles include Yukawa and scalar couplings to exotic light degrees of freedom.

\paragraph{The mass of the DM state. } Equation~\eqref{eq:FractionalAsymmetry} does not alone determine the relic DM abundance. In the presence of an asymmetry, reproducing the observed DM density yields a prediction for the DM mass, as per
\begin{equation}
\frac{m{_{_{\rm DM}}}}{m_p}  = 
\frac{\Omega{_{_{\rm DM}}}}{\Omega_{\rm _{VM}}} \:
\frac{\eta (B_{\rm V})} {\eta (B_{\rm D}) / q{_{_{\rm DM}}}} \:
\frac{1-r_\infty}{1+r_\infty} 
\ ,
\label{eq:ADM mass}
\end{equation}
where $m_{_\text{DM}}, \ q_{_\text{DM}}$ is the mass and the dark-baryonic charge of the DM state, $m_p$ is the proton mass, and $\eta(B_V), \ \eta(B_D)$ are the ordinary and dark baryonic charge-to-entropy ratios. The DM mass is of course critical in assessing the direct and indirect detection prospects of DM models.

In baryon-symmetric models, corresponding to the symmetry structure (i) described above, $\eta(B_V) = a_s \eta(B_D)$, where $a_s \simeq 0.35$ or 1, depending on whether the generation of the asymmetries happened before or after the electroweak phase transition, respectively~\cite{Harvey:1990qw}. Then, from Eq.~\eqref{eq:ADM mass} and for $r_\infty \to 0$, the DM mass has to be 
$m_{_\text{DM}} \simeq q_{_\text{DM}} \times (1.6 - 5) \text{ GeV}$. 
If effects similar to the electroweak sphalerons are operative in the dark sector after the asymmetry generation has taken place, the above prediction may be modified by a factor of a few.

If the relation between the visible and dark asymmetries is established via chemical equilibrium, as in  the symmetry structure (ii) described above, then the DM mass required to reproduce the correct DM abundance depends on the details of the chemical equilibrium. Chemical equilibrium tends to keep the chemical potentials of the visible and dark sector particles at the same magnitude. 
If it ceases when at least some of the visible and the dark baryonic species are still relativistic, then the corresponding number densities, and thus asymmetries, are of the same magnitude. This implies that DM should be in the GeV range (for $q_{_\text{DM}} \sim {\cal O}(1)$). However, if the  chemical decoupling of the two sectors occurs when DM is non-relativistic, while the SM quarks and leptons are still relativistic, the number density of the dark species is Boltzmann suppressed, and $n(B_D)/n(B_V) \sim \exp (-m_{_\text{DM}}/T)$. In this case a much larger DM mass is required in order to compensate for the thermal suppression of the DM number density. The exact value depends of course on the details of the chemical equilibrium, but typically it is expected to be $m_{_\text{DM}} \sim \text{ TeV}$~\cite{Barr:1990ca,Buckley:2010ui}.

The above estimates for the DM mass are modified if the dark sector involves additional dynamics which allow the DM particles to form bound states. This is the case \eg~in mirror DM models, where dark baryons are bound in heavy mirror atoms due to the mirror nuclear interactions~\cite{Foot:2003iv,Foot:2010hu,Foot:2011pi,Foot:2013msa}, and in $Q$-ball DM scenarios with $Q$-balls being bound states of scalar fields carrying a conserved global charge, and are typically stable for masses $m_Q > 10^{12} \text{ GeV}$~\cite{Kusenko:1997ad,Kusenko:1997zq,Kusenko:1997si}.

The above shows that the possibilities for the DM mass in the ADM scenario span a large range of values. However, in most models, the DM mass is predicted to be in the range (1-15)~GeV, as a consequence of the similar visible and dark baryonic asymmetries. The positive signals that have been claimed by a few DM direct detection
experiments~\cite{Bernabei:2008yi,Bernabei:2010mq,Aalseth:2010vx,Aalseth:2011wp,Angloher:2011uu,Agnese:2013rvf}, all pointing in this mass region, lend thus support to the idea that the VM and DM number densities are similar and were related via some dynamics that took place in the early universe. Nevertheless, a theoretical justification for the DM mass scale and its relation to the QCD scale is absent in most ADM models, with mirror DM (discussed later) being an exception. Another approach which also assumes a confining group in the dark sector producing the DM mass scale, and relates the infrared fixed points of 
this and the ordinary QCD, has been recently explored in Ref.~\cite{Bai:2013xga}.

\subsubsection{Phenomenology}

\paragraph{Extra radiation. } The efficient annihilation of the symmetric part of DM raises the possibility of the existence of dark radiation in the universe. If the dark sector involves stable light species, as often appears in ADM models, they will contribute to the relativistic energy density of the universe and potentially affect BBN and CMB. 
Radiation present in our universe that exceeds what can be accounted for by photons and the three known neutrino species is customarily quantified in terms of extra neutrino species, $\delta N_\text{eff}$, as per
\begin{equation}
\delta \rho = \frac{7 \pi^2}{120} \left( \frac{4}{11} \right)^{4/3} 
\delta N_\text{eff} \: T_{_{\rm V}}^4 \ ,
\label{eq:extra rad}
\end{equation}
where $\delta \rho$ is the extra relativistic energy density and $T_{_{\rm V}}$ is the ordinary photon temperature. Current observations cannot rule out $\delta N_\text{eff}$ being a substantial fraction of 1. 
If the dark sector contains new stable relativistic species, their contribution to the energy density of the universe is $\rho_{_{\rm D}} = g_{_{\rm D}} (\pi^2/30) T_{_{\rm D}}^4$, where $g_{_{\rm D}}$ is the temperature-dependent effective number of relativistic degrees of freedom in the dark sector. $T_{_{\rm D}}$ is the dark-sector temperature which is generically different from $T_{_{\rm V}}$. How much $T_{_{\rm D}}$ can deviate from $T_{_{\rm D}}$ depends on the relative complexity of the dark and the visible sectors. Constraining the amount of non-standard radiation thus provides information not only about the DM itself, but also about the hidden sector it may inhabit.

\paragraph{Structure formation and galactic dynamics. }

The ADM scenario can accommodate a range of possibilities for the properties of DM which affect its gravitational clustering, such as the nature and strength of the DM self-interaction, and the coupling strength of DM to (dark) radiation.
Asymmetric DM can behave as collisionless CDM with early kinetic decoupling from relativistic species; 
indeed, there are many ways to approach this limit in a variety of ADM constructions. 
That it is sufficient for the ADM annihilation cross-section to be only slightly larger than the canonical value for symmetric thermal-relic DM, as discussed around Eq.~\eqref{eq:sigma ann}, testifies to the fact that the properties of ADM  may be unobservably different from the properties of WIMP DM.
However, the ADM annihilation cross-section can be anywhere in the continuum described by Eq.~\eqref{eq:sigma ann}. This freedom and the possibility of ADM being part of a potentially rich hidden sector with its own (gauge) interactions and/or light species, allow DM to possess non-standard properties with observable implications for structure formation.

Asymmetric DM can comfortably have sizable self-interactions which can affect galactic dynamics and resolve the structure-formation problems of collisionless CDM (see Sec.~\ref{sec:self-interDM} for more details on the appropriate range of cross-sections and nature of self-interactions). To compare the plausibility of significant self-interaction in the ADM scenario with that in the WIMP paradigm, we may discern two cases: 
First, consider self-interactions mediated by a species lighter than DM itself. In this case, the same coupling which gives rise to the DM self-scattering, also contributes to the DM annihilation. Because the annihilation cross-section for symmetric DM has a specific value, while for ADM it can span an unbounded continuum, the parameter space which yields the desired DM self-interaction is obviously greater in the ADM scenario than in the symmetric DM scenario. It should be noted that interactions mediated by light particles can potentially behave as long-range and yield velocity-dependent cross-sections, depending on the interplay of the various parameters involved. Long-range interactions are particularly suitable for resolving the structure formation problems of collisionless CDM without spoiling its successful predictions~\cite{Vogelsberger:2012sa}. Atomic DM~\cite{CyrRacine:2012fz} and mirror DM~\cite{Foot:2013lxa,Foot:2013vna,Foot:2013uxa,Foot:2004wz}
are examples of ADM models which feature long-range interactions mediated by a massless vector boson\footnote{Note though that the dynamics of mirror DM is more complicated, due to dissipation and heating, and cannot be captured by simply considering the DM self-scattering~\cite{Foot:2013lxa,Foot:2013vna,Foot:2013uxa,Foot:2004wz}. See Sec.~\ref{sec:mirrorDM} for more details.} . Yukawa interactions can also result in $v$-dependent DM self-scattering~\cite{Feng:2009hw,Loeb:2010gj,Tulin:2013teo}. The characteristics of the DM self-interaction for these examples are described in more detail in Secs.~\ref{sec:mirrorDM} and \ref{sec:self-interDM}.
Second, consider DM self-interactions involving heavier fields, such that the relevant couplings do not contribute to DM annihilation. The DM self-scattering can then be described by an effective operator (for scalar DM such couplings also feed into the renormalizable quartic coupling). The resulting DM self-scattering cross-section can be sizable within the perturbativity limit of the dimensionless coupling involved, only if the scale of these operators is sufficiently low, thus implying that DM has to be light. The ADM scenario motivates a lower DM mass scale than the WIMP scenario, and can thus accommodate sizable contact-type DM self-interactions more comfortably, even if such interactions are not associated with a contribution to the DM annihilation cross-section.

The plausible direct coupling of ADM to light (or massless) degrees of freedom has another important implication. In the early universe, ADM must have been kinematically coupled to a thermal bath of dark radiation. If its kinetic decoupling from dark radiation occurred late, it may have affected the growth of matter-density perturbations at small (\ie~dwarf-galaxy) scales.  Late kinetic decoupling is more likely in the ADM scenario than in the symmetric WIMP scenario, because ADM must annihilate more efficiently and is thus expected to interact more strongly with radiation.\footnote{This may be the case provided that the temperature of the dark plasma is not very different from the temperature of the SM radiation at the time of DM kinetic decoupling. Otherwise, the kinetic decoupling of ADM may occur earlier than what is usually estimated for the kinetic decoupling of  symmetric WIMPs from SM particles.}
During the decoupling epoch both the damping of dark baryon acoustic oscillations~\cite{Loeb:2005pm,Bertschinger:2006nq} and dark-radiation diffusion (Silk damping~\cite{Silk:1967kq}) can reduce the amplitude of sub-horizon perturbations. As a result, the formation of structure at scales smaller than the damping horizon gets suppressed~\cite{Foot:2012ai,CyrRacine:2012fz,Feng:2009mn}.

\paragraph{Direct detection. }

Potential channels for DM direct-detection in ADM models include exchange of new gauge bosons or scalar particles. 
In models which feature the symmetry structure (i) described earlier, the 
always conserved particle number $B_\text{con}$, may arise as the global remnant of an Abelian gauge symmetry under which both the ordinary and the dark baryons are charged. In this case, a plausible channel for direct detection is the exchange of a $Z'$ (which could have the same couplings with SM particles as a $Z'_{B-L}$). 
Another well-motivated possibility is nucleon-DM interaction via a dark photon which mixes kinetically with hypercharge, as in Eq.~\eqref{eq:kinetic mixing}. Nucleon-DM scattering can also be mediated by a scalar particle which couples to DM and mass-mixes with the SM Higgs. 

In all these cases, the DM-nucleon interaction can be described by a Yukawa potential $V(r) = \alpha \exp (-m_\phi r) / r$, where $m_\phi$ is the mass of the vector or scalar mediator. This gives rise to a DM-nucleon differential cross-section 
with two very different limiting regimes
\begin{equation}
\frac{d\sigma(v, E_R)}{d E_R} =
\left \{
\begin{alignedat}{5}
&\frac{8\pi \alpha^2 \, m_N}{m_\phi^4} \, \frac{F^2(E_R)}{v^2}  \ , &
\qquad 
& m_\phi^2 \gg 2 m_N E_R : \text{ short-range interaction}&
\\
&\frac{2 \pi \alpha^2 }{m_N} {} \, \frac{F^2(E_R)}{v^2 E_R^2 }  \ , &
\qquad 
& m_\phi^2 \ll 2 m_N E_R : \text{ long-range interaction} \ , &
\end{alignedat}
\right.
\label{eq:sigmaDD regimes}
\end{equation}
where $m_N$, $E_R$ are the mass and the recoil energy of the nucleus, $F(E_R)$ is the nuclear form factor and $v$ is the speed of the DM particle. When the momentum transfer  $q^2 = 2 m_N E_R$ is $q^2 \ll m_\phi^2$, the DM-nucleon interaction is contact-type, and the usual interpretation of the direct-detection experiments, which assumes WIMP DM, applies. If $q^2 \gg m_\phi^2$, which includes the case of a massless mediator, the DM-nucleon interaction is long range. The $E_R^{-2}$ dependence of the differential cross-section changes the interpretation of the direct-detection results and has the potential to bring the various direct-detection experiments into better agreement~\cite{Foot:2003iv,Foot:2010hu,Foot:2011pi,Foot:2013msa,Foot:2012cs,Fornengo:2011sz}.
For typical targets of mass $m_N \sim 100 \text{ GeV}$ and nuclear recoil energies around $E_R \sim 10 \text{ keV}$, interactions manifest as long-range if $m_\phi \lesssim 50 \text{ MeV}$.

In the short range regime, exchange of a $Z'_{B-L}$ or a massive dark photon with kinetic mixing with hypercharge, give spin-independent DM-nucleon scattering cross-sections
\begin{eqnarray}
\sigma_{B-L}^{\rm SI}   &\approx&   
10^{-46} \text{ cm}^2  \: \times \:  q_{_\text{DM}}^2
\left( \frac{g_{_{B-L}}}{0.1} \right)^4 
\left( \frac{3 \text{ TeV}}{M_{B-L}} \right)^4,
\label{eq:sigmaDD_B-L}
\\
\sigma_D^{\rm SI}    &\approx&   
10^{-40} \text{ cm}^2
\left( \frac{\epsilon}{10^{-4}} \right)^2      
\left( \frac{g_{_D}}{0.1} \right)^2     
\left( \frac{1 \text{ GeV}}{M_D} \right)^4 .
\label{eq:sigmaDD_dark}
\end{eqnarray}
The above cross-sections have been evaluated for $m_\text{DM} = 5 \text{ GeV}$. The exchange of a massive dark photon can account for the low-mass signals identified by DAMA~\cite{Bernabei:2008yi,Bernabei:2010mq}, CoGeNT~\cite{Aalseth:2010vx,Aalseth:2011wp}, CRESST~\cite{Angloher:2011uu} and CDMS~\cite{Agnese:2013rvf}, but it can also vary by a few orders of magnitude and satisfy the limits on short-range DM-nucleon interactions from XENON~\cite{Aprile:2011hi,Angle:2011th,Baudis:2012zs,Essig:2012yx}.  

The long-range regime appears commonly in ADM models, which can accommodate sizable DM couplings to light mediators without spoiling the DM relic abundance. Because in this regime  $d\sigma/dE_R \propto  E_R^{-2}$, experiments with low-energy thresholds, such as DAMA and CoGeNT, are more sensitive than experiments with higher energy thresholds, such as XENON100.  The total DM-nucleon scattering cross-section is velocity-dependent  ($\sigma_{n\chi} \propto v^{-4}$ for $m_\phi \to 0$), obviating the fact that the interpretation of the direct-detection results in terms of the fundamental particle-physics parameters involved is different than in the case of short-range interactions~\cite{Foot:2003iv,Foot:2010hu,Foot:2011pi,Foot:2013msa,Foot:2012cs,Fornengo:2011sz}. 
It has been shown that the existing signal regions from different direct-detection experiments can be brought in agreement if the DM-nucleon scattering is long-range, mediated by either a massless~\cite{Foot:2003iv,Foot:2010hu,Foot:2011pi,Foot:2013msa,Foot:2012cs} or a light particle~\cite{Fornengo:2011sz}, with preferred DM and mediator masses around $m_\text{DM} \sim 10\text{ GeV}$ and $m_\phi \sim 10\text{ MeV}$  (see also Sec.~\ref{sec:mirrorDM}).
Compatibility with the bounds from XENON100 is may be achieved when reasonable estimates for the systematic uncertainties are considered~\cite{Foot:2013msa,Foot:2012cs}.

Another factor which contributes to better compatibility among direct-detection experiments is a modified velocity dispersion. This  arises naturally within the mirror DM scenario, as a result of DM being multi-component (albeit with each component made of the same fundamental constituents) and self-interacting. In such a scenario, every component has velocity dispersion which depends on its mass, $v_i \propto m_i^{-1/2}$ (see Sec.~\ref{sec:mirrorDM} for more details). The same features could plausibly occur in other similar to mirror DM constructions within the ADM framework.

\paragraph{Indirect detection. }

The ADM scenario does not predict significant annihilation signals. The rate of particle-antiparticle annihilations in today's universe is exponentially suppressed with increasing annihilation cross-section, due to the absence of antiparticles today (c.f. Eq.~\eqref{eq:FractionalAsymmetry}). Self-annihilations --if at all possible-- are suppressed by the lowest scale at which $B_\text{D}$ violation occurs. If it is to relate the visible and the dark baryonic asymmetries, this has to be the same as the lowest scale at which $B_\text{V}$ or $(B-L)_\text{V}$ occur. The latter  are constrained by various considerations to be rather high.

On the other hand, the joint violation of $B_\text{D}$ and $(B-L)_\text{V}$ can potentially allow DM co-annihilation with SM fermions, provided that the corresponding interactions do not preserve any discrete subgroup of  $(B-L)_\text{V}$ or $B_\text{D}/q_{_\text{DM}}$, where $q_{_\text{DM}}$ is the dark baryonic charge of DM. 
It has been proposed that DM-VM coannihilation can be potentially observed as ``induced nucleon decay" in terrestrial nucleon-decay experiments. Incident DM particles can co-annihilate with nucleons, producing detectable mesons of higher energy than expected in the case of spontaneous proton decay, thus distinguishing between the two processes~\cite{Davoudiasl:2011fj}. 

The same couplings could cause DM decay into SM particles (and possibly dark radiation), provided that this is kinematically allowed. An interesting feature of ADM decays in SM degrees of freedom is that the decay products include asymmetric amounts of SM particles and antiparticles. 
If the DM couplings to SM particles  are flavor-dependent, the final decay products will exhibit overall an energy-dependent charge asymmetry~\cite{Chang:2011xn,Masina:2011hu}, with charge neutrality being of course ensured by oppositely-charged decay products produced with different energies at different stages of the decay chain.  This can be a powerful signature of flavor-violating decaying ADM, and could potentially explain the tension between the recent AMS-02 measurements of the positron-fraction spectrum and the Fermi-LAT measurements of the total electron+positron flux~\cite{Masina:2013yea,Feng:2013vva}. 

Another process which can yield interesting indirect detection signals is bound-state formation in the galaxies today~\cite{Pearce:2013ola}. Asymmetric DM can have significant self-interactions; if attractive, these interactions imply the existence of bound states in a large range of the parameter space. In the absence of annihilations between DM particles, the bound states are stable rather than short-lived. Bound-state formation is invariably accompanied by emission of radiation. Being an exothermic process, it is favored to occur over elastic scattering when DM collisions take place. The radiation emitted from formation of bound states in the galactic halos can yield detectable signals~\cite{Pearce:2013ola}.

\paragraph{Capture in stars. }

Asymmetric DM captured in stellar objects accumulates over time, without its density being capped by annihilations. The dense accumulation of ADM in stars has a number of consequences which have allowed for constraints to be placed on the ADM parameter space.

Dark-matter particles captured in a star quickly thermalize via their collisions with nucleons and sink in the core the star, within some thermal radius which depends on the temperature and the density of the star and the DM mass. If the mean free path of the DM particles, which depends on their scattering cross-section with nucleons (and on their self-scattering cross-section, if self-interactions are important), is larger than the thermal radius, DM causes non-local energy transport from the innermost part of the star to its outer regions. Non-local energy transport can affect the thermal conductivity, the sound speed, the convection zone and the oscillation modes of stars. Precision asteroseismology, available for the Sun and other systems such as the binary $\alpha$~Cen~B, constrains the allowed variation of these properties, and thus places bounds on the ADM parameters. Another promising probe is the solar neutrino fluxes which depend very sensitively on the temperature and the density profile of the 
inner regions of the Sun.
Current analyses of the effect of ADM capture in the Sun and other main-sequence stars disfavor the mass region $m_{_\text{DM}} \sim (5 - 20) \text{ GeV}$ for ADM with short-range interactions with nucleons at spin-dependent scattering cross-sections in the range $\sigma_{n\chi} \sim (10^{-36} - 10^{-33}) \text{ cm}^2$ and no self-interactions~\cite{Gelmini:1986zz,Frandsen:2010yj,Cumberbatch:2010hh,Taoso:2010tg,Lopes:2012af}.

For smaller DM-nucleon scattering cross-sections, DM can be captured efficiently in compact objects, such as white dwarfs and neutron stars. Asymmetric DM that accumulates in the center of compact objects  can condense collectively, and eventually reach the critical density for gravitational collapse. A mini black hole can form and potentially consume the host star. The observation of old compact stars can thus constrain the properties of ADM.

For fermionic ADM with short-range spin-dependent interactions with nucleons and no (repulsive) self-interactions,  the constraints  from observations of old neutron stars in the globular clusters, assuming local DM density $\rho_{_\text{DM}} \sim (10^3 - 10^4) \text{ GeV cm}^{-3}$, are competitive with the constraints from direct-detection experiments for DM masses $m_{_\text{DM}} \gtrsim \text{TeV}$~\cite{Kouvaris:2010jy}. Attractive self-interactions lower the number of particles necessary for gravitational collapse. For fermionic ADM interacting via an attractive Yukawa potential, observations of nearby old pulsars  place limits on the strength of the interaction and the screening scale, which extend beyond the limits from the Bullet Cluster~\cite{Kouvaris:2011gb}.

Bosonic matter in its ground state is supported by the uncertainty principle, with the critical mass for gravitational collapse being significantly lower than in the case of fermionic matter. For fundamental bosonic ADM with vanishing self-interaction, observations of old neutron stars exclude the mass range $m_{_\text{DM}} \approx 1 \text{ MeV} -16 \text{ GeV}$ for DM-nucleon cross-sections $\sigma_{n\chi} \gtrsim 10^{-43} \text{ cm}^2 $ (with a more narrow mass range excluded at smaller $\sigma_{n\chi}$)~\cite{Kouvaris:2011fi,McDermott:2011jp}. However for fundamental bosons, the quartic self-interaction term is not protected by any unitary symmetry~\cite{Bramante:2013hn,Bell:2013xk}, and is inevitably generated by 
the interaction of DM with nucleons~\cite{Bell:2013xk}. The latter is necessary for the capture of DM in the star. A tiny repulsive self-interaction is sufficient to offset the limits on bosonic ADM~\cite{Kouvaris:2011fi,Bramante:2013hn,Bell:2013xk}, leaving most bosonic ADM models largely unconstrained. (No constraints currently exist for bosonic ADM with attractive self-interactions). 
However, the quartic self-interaction might be absent in supersymmetric models, where the scalar potential typically contains a large number of flat directions~\cite{Bell:2013xk}. The bounds then apply to models in which ADM corresponds to a flat direction in a supersymmetric theory, as for example  in the model of Ref.~\cite{Mitropoulos:2013fla}, albeit the exclusion region is shrunk due to supersymmetry breaking~\cite{Bell:2013xk}.

\subsubsection{Asymmetric Dark Matter: Summary and outlook}

The ADM scenario is motivated by the supposition that both the microphysics and the cosmological history of VM and DM are closely related, with current observations supporting this hypothesis.

When assessing the phenomenological implications of ADM, it is useful to start with the question: Does the cosmology and the direct and indirect signatures of ADM \emph{have} to be different from those of symmetric thermal-relic DM? The answer to this is no. Asymmetric DM can resemble closely the standard WIMP DM, to which it is in fact related; indeed, ADM encompasses all viable thermal-relic DM models with annihilation cross-section larger than that for symmetric thermal-relic DM. The latter is, in this sense, a limiting case of a continuum spanned by ADM.
But a more interesting question is this: Can the observable properties of ADM differ from those of symmetric WIMP DM, and is this possibility of interest? The answer to this is yes. Asymmetric DM can inhabit a potentially rich sector with rich phenomenology. Finding a non-standard property of DM would be a huge breakthrough in understanding its nature, and it is thus essential that we understand the range of possibilities that ADM (and other DM models) provide. In fact, 
there are currently observations which hint that the DM properties deviate from those of the WIMP paradigm, although it is not yet possible to conclude this with certainty. The question then morphs into this: How different from standard \emph{should} the DM properties be within the quasi-continuum of possibilities that exist?

Due to the potential richness of its dark sector and the continuous range of annihilation cross-sections it may have, ADM  can be a prototype for self-interacting DM, which currently appears to explain the observed galactic structure more adequately than collisionless CDM. Yet, the possibilities for the nature of the ADM self-interaction and the implications for structure formation have not been fully investigated, neither theoretically nor numerically. The ADM couplings with itself and other dark-sector particles and with VM have been shown to give potentially observable indirect detection signatures, \eg~via coannihilations with VM, decay or bound-state formation. These deserve further investigation. Considering the effects of ADM capture in stars has been a powerful method in constraining ADM interactions, which could be potentially extended to a larger variety of ADM models, \eg~models with long-range self- and DM-nucleon interactions. On the theoretical side, perhaps what stands as the most important 
open question in the ADM scenario is the origin of the DM mass-scale and its relation to the QCD scale, for which relatively few suggestions currently exist. 

It is often asked: What would be a smoking-gun signature of ADM? The answer is, not surprisingly, complex. It is rather implausible that the nature of DM, be it standard WIMPs or anything else, can be fully identified by a single piece of evidence or a single experiment. However, if a set of signals can be collectively and successfully attributed only to DM with properties which imply that the DM annihilation cross-section is larger than the canonical value for symmetric DM, then this would constitute very strong evidence for ADM. Interestingly, an example pointing to this direction already exists: 
Direct-detection experiments, appearing to be in conflict when interpreted within the WIMP paradigm, can be brought in better agreement if long-range DM-nucleon interactions, mediated via a massless or light boson, are considered. This possible explanation involves two new couplings: the DM coupling and the nucleon coupling to the light mediator. 
Within this interpretation, the  signal strengths imply that DM would have to annihilate more efficiently than symmetric DM, via the DM coupling to the light mediator (with the nucleon-mediator coupling being bounded from above by other experiments, thus forcing a lower limit on the DM-mediator coupling for the purpose of explaining the observed direct-detection signals). 
Although certainly not yet conclusive, this piece of evidence shows why we would do well to explore a  large range of possibilities for the DM physics, and why identifying the nature of DM requires in fact input from many and diverse experiments.

\subsection{Axions} 
Axions, which arise as a consequence of introducing a global chiral symmetry to resolve the strong CP problem in QCD, are one of the best motivated candidates for the Universe’s dark matter. 

There is an serious problem in the Standard Model called the strong CP problem.
The strong CP problem originates because the structure of the QCD vacuum induces an extra term in the QCD Lagrangian of the standard theory involving the
gluon density $ G_{a\mu\nu}\tilde{G}_a^{\mu\nu}$,  
where $ G_{a\mu\nu}$ is the gluon field strength and $ \tilde{G}_a^{\mu\nu}= \frac{1}{2} \epsilon^{\mu\nu\alpha\beta} G_{a\alpha\beta}$ is its dual,
\begin{equation} 
L_{\theta}=\theta \frac{\alpha_s}{8\pi}G_{a\mu\nu}\tilde{G}_a^{mu\nu}
\end{equation}                 
This term violates P and T, but conserves C, and thus can produce a neutron electric dipole moment of order $d_n \simeq \frac{ e m_q}{M_n^2} \theta$.
It's also the case that this term is necessary to explain the masses of the light mesons (the "$\eta$-mass problem). Hence, there are strong reasons
to believe this term is part of the Standard Model.  The observable quantity in the Standard Model is a combination of the phase from QCD vacuum and a phase coming from the quark Yukawa couplings.  The sum of these two independent contributions could be expected to be a number of order one. 
However, the strong bound on the 
neutron edm, $d_n<1.1 \times10^{-26}$ ecm~\cite{Baker:2006ts}. requires the angle $\theta$  to be very small: $\theta < 10^{-10}$.
Why this should be so is the strong CP problem~\cite{Cheng:1987gp}.

The only viable solution of this problem, proposed by Peccei and Quinn in 1977~\cite{Peccei:1977hh,Peccei:1977ur} introduces a global chiral symmetry which is spontaneously broken. This, so-called, $U(1)_{PQ}$ symmetry dynamically drives the parameter  $\theta$ to zero. Because $U(1)_{PQ}$ is spontaneously broken, a Nambu--Goldstone boson
arises in the theory, and it is this particle which is called the
``PQ axion'' or ``QCD axion''~\cite{Weinberg:1977ma,Wilczek:1977pj}. In effect, as the result of the extra $U(1)_{PQ}$ symmetry, the static CP-violating interaction in QCD characterized by the parameter $\theta$ gets replaced by a CP conserving interaction of the axion field $a(x)$ with the gluon density. That is,
\begin{equation}
\theta \to \frac{a(x)}{f_a},
\end{equation}
where $f_a$ is the scale associated with the spontaneous breaking of the U(1)$_{PQ}$.

All properties of the axion (mass, couplings and lifetimes) depend on the scale $f_a$~\cite{Kim:2008hd,Beringer:1900zz}. For instance, the interactions of the axion with other particles in the theory are inversely proportional to $f_a$. The coupling of the axion to two photons, which is similar to its couplings to two gluons,
is also proportional to $1/f_a$. Finally, because the $U(1)_{PQ}$ symmetry has a chiral anomaly,
the axion acquires a small mass which is also inversely proportional to $f_a$. The strong CP problem is resolved, however, irrespective of the value of this scale parameter.
This is similar in spirit to Higgs physics.

In the model considered originally by Peccei and Quinn (the ``classic axion''), the scale $f_a$ was associated with scale of the breaking of the electroweak interactions. In this case, axions have masses in the keV range, but such axions have have relatively strong interactions and have been ruled out experimentally. However, if there are fields in the theory which carry a PQ charge, but are neutral under the electroweak interactions, the scale $f_a$ can be much greater that the electroweak scale. Two benchmark models are often considered. In the, so-called, DFSZ models \cite{Dine:1981rt,Zhitnitsky:1980tq} an $SU(2) \times U(1)$ singlet field carrying PQ charge is added to the model proposed by Peccei and Quinn. This is very attractive in that the axion coupling to two photons is simply related to properties of the known particles. In the, so-called, KSVZ models~\cite{Kim:1979if,Shifman:1979if} on the other hand, only a heavy quark and an $SU(2) \times U(1)$ singlet field carry PQ charge. This axion has no 
couplings to leptons. In all these models, the resulting axions that would make up dark matter are very long lived, very light and very weakly coupled to normal matter and radiation. Such axions have therefore been dubbed “invisible” axions. This notwithstanding, if they exist, such invisible axions  have important astrophysical and cosmological consequences. Further such axions would be detectable in the newest generation of exquisitely sensitive axion searches.

Astrophysics gives a lower bound on $f_a$ since axion emission, both through  its coupling to two photons (Primakoff process) and as a result of axion bremsstrahlung, causes energy losses in stars and other astrophysical objects~\cite{Raffelt:1996wa}, for example, evolution of stars in the 
8--10~$M_\odot$ mass window~\cite{Friedland:2012hj}. Since these losses are proportional to $1/f_a$, the $U(1)_{PQ}$ scale has to be large enough to prevent altering known astrophysical processes. Typically, by looking at star cooling rates one deduces that $f_a> 6 \times 10^6$ GeV. Stronger, but perhaps more uncertain,  bounds come from an analysis of the neutrino pulse from  SN1987a giving $f_a  > 6\times 10^9$~Gev~\cite{Raffelt:1996wa}. 

Cosmology, on the other hand, gives an upper bound for the $f_a$  scale due to the ``misalignment mechanism'' of axion production in the early universe.
The physics that gives rise to this bound is simple to 
understand~\cite{Preskill:1982cy,Abbott:1982af,Dine:1982ah}. At a temperature $ T \sim f_a$  when the Universe goes through the phase transition which leads to the breakdown of $U(1)_{PQ}$, the QCD chiral anomaly is ineffective. Thus, the expectation value of the axion field $\langle a \rangle_{\rm init}$ is arbitrary. Eventually, when the Universe cools to temperatures of order of the dynamical scale of QCD ($ T \sim \Lambda_{QCD}$ the axion gets a mass and the axion vacuum expectation value is driven to its CP conserving value $\langle a \rangle \to – f_a \theta$. The resulting coherent zero momentum axion oscillations towards this minimum contribute to the Universe’s energy density and act as cold dark matter. This contribution to the dark matter of the Universe is proportional to the initial axion expectation value squared $\langle a \rangle_{\rm init}^2$, divided by the scale $f_a$. One sees that, if $\langle a \rangle_{\rm init} \sim f_a$, then a bound on the dark matter of 
the Universe provides an upper bound for $f_a$. Using the present bounds on the density of cold dark matter and $\langle a \rangle_{\rm init} \sim f_a$ gives an upper bound $f_a < 3 \times 10^{11}$ GeV.

In some models, inflation allows one to evade the upper bounds on $f_a$ and the corresponding lower bound on the mass of the axion~\cite{Linde:1987bx}.  
Since inflation allows a broad range of possible values of the axion field in parts of the universe that are separated by superhorizon distances, the 
anthropic selection criteria apply to the value of the field in the observable universe~\cite{Tegmark:2005dy,Wilczek:2013lra}.  This broadens the range of acceptable masses for axion dark matter.

\noindent\leftline{\bf Axion Searches}

Searches for axions can be organized into several groupings
One grouping is that of Peccei-Quinn (PQ) axions versus non-PQ axions.
As discussed earlier, PQ axions (or ``QCD axions'') have a relatively tight relation
between the axion mass and the axion couplings to normal matter
and radiation. Were these PQ axions to be the dominant component of
dark matter, their masses would be in the approximate range
1$\mu$eV to 100$\mu$eV, representing two decades of allowed axion mass.
Recall such axions were dubbed ``invisible axions'' due to
their exquisitely feeble couplings to normal matter and radiation.
These PQ axions especially are the target of the ``RF-Cavity'' axion searches,
which will explore these two decades in the near future at
high sensitivity. An additional
approximate mass decade 100$\mu$eV to 1000$\mu$eV is
mainly unexplored at the sensitivity level of PQ-type axions, though this
decade would be disfavored should PQ axions form the dominant
component of the dark matter; such axions would not be produced
in enough abundance to be dark matter seen today.
Exploring this upper decade
of axion masses with RF-cavities requires advances in detector technology, as photons
from such axions are upwards of many THz in frequency. However, axions
in this third allowed decade may be detectable by solar axion searches.
Since dark-matter may not be axions of the PQ type, it is prudent
to have a science program that encompasses non-PQ
type axions. These non PQ-type axions weaken or remove the relation between the axion
mass and its couplings to normal matter and radiation. This opens up
an enormous search space of axion masses and couplings, but at the
same time a negative result from a particular search for non-PQ axions will
not in general rule out a particular class of non-PQ axion models.
Some of these non-PQ axions could make up the dark matter, others
are interesting from the standpoint of extremely weakly interacting
particles, but do not necessarily speak to the dark matter problem. Some
of these non-PQ axion searches overlap with the charge of the Intensity Frontier
working groups' reports and are discussed in more detail there.

Another organizing principle of axion searches is that of terrestrial
versus astrophysical axion searches. Terrestrial searches may depend
on the axions being remnants from the Big Bang (in the case of
RF-Cavity searches), from the Sun (in the case of the ``helioscope''
type detectors), or through the axions' effects as virtual particles
(in the case of laser type detectors, the case of short-distance
gravity detectors, or the recent ideas of looking
for time-varying T violating effects in molecules or electron spins).
The astrophysical searches for axions look for the effects of axions
on astrophysical objects or in the propagation of photons through
intergalactic space. In some cases, axions produced in the
astrophysical object alter the non-axion energy production. This
would, for example, change the evolutionary age of the Sun compared
to its chronological age. It would also effect the Suns seismology.
Axion production would have also altered the arrival-time distribution
of the neutrinos detected from SN1987A. Axions would also alter the
evolution red giant evolution and thereby alter the population
of horizontal-branch stars. Axions could also effect the cooling of
white dwarf stars and thereby alter the the white dwarf luminosity
distribution. In other cases axions would allow photons to have
unusually long absorption lengths due to their mixing with axions
in the presence of intergalactic magnetic fields.

At present, the search window for QCD axions is established by the SN1987A neutrino signal
(generally excluding PQ type axions with masses greater than around 1000~$\mu$eV)
and the theoretical argument that too-light axions would produce too much dark matter.
The latter bound (generally excluding axions with masses less than around 
1$\mu$eV)can be evaded in the context of inflationary 
cosmology~\cite{Wilczek:2013lra,Tegmark:2005dy,Linde:1987bx}.  
Interestingly, the RF cavity experiments
are now sensitive PQ axion couplings in this window where
such axions would form a majority component of the dark matter.
For non PQ type axions, the search space of mass and couplings
is large and much of it remains ripe for exploration. 

In the following discussions, the main grouping will be that of
terrestrial versus astrophysical searches. With this grouping,
PQ type and non-PQ type axions are typically searched for in
the same apparatus.

\noindent\leftline{\underline Terrestrial Searches}

Recall that the search for PQ axions in the 1--100~$\mu$eV mass range is
extremely difficult due to their extraordinarily weak couplings.
For instance, an axion of this mass has a lifetime in the neighborhood
of $10^{50}$ years before it decays into two photons.
Such minuscule couplings make neutrino couplings by comparison seem very
strong. Amazingly, terrestrial experiments are sensitive too such weakly-interacting
particles.

A. RF-Cavity Searches

Although these searches are difficult, advances in the detection of low levels of
electromagnetic radiation open the way for searches
for these ``invisible axions'' through their decay
into microwave photons. In these experiments, a volume
of space is threaded by a multi-tesla static magnetic field.
Nearby Milky Way halo axions, and such axions have a de Broglie
wavelength much longer than the experiment dimensions,
scatter off this potential and thereby
convert into a single microwave photon carrying almost the
entire mass plus kinetic energy of the incident axion.
The magnet absorbs a very small momentum kick in this process.
To increase the conversion rate, the conversion volume is
encased in a high-Q microwave cavity tuned to the frequency
of the outgoing photon, thereby increasing the rate by the
factor Q. The power released in the cavity from such axion
to photon conversions is small, perhaps in the neighborhood
of $10^{-21}$ to $10^{-23}$ watts. However, the most recent
receivers used in these experiments have demonstrated sensitivity
of better than a yoctowatt ($10^{-24}$ watts). Such low noise
receivers operate at the noise floor allowed by the
standard quantum limit~\cite{Sikivie:1983ip}.

\begin{figure}[!htb]
  \includegraphics[width=\textwidth]{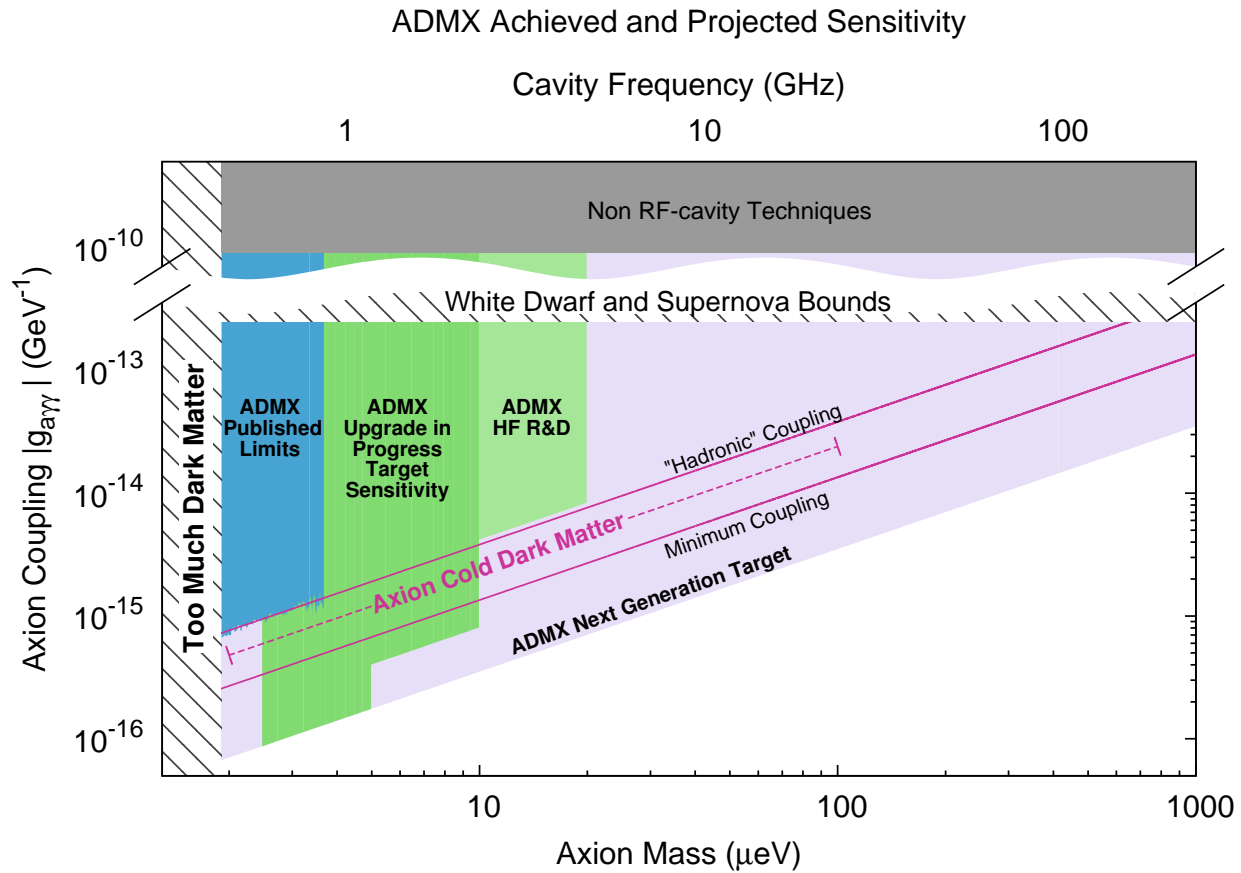}
\caption{The search reach of the ADMX RF-cavity experiments over the next 3 years.
 The first decade of allowed axion mass will be explored at ``definitive'' sensitivity
 to QCD axions over the next year. The middle decade will be explored at over the following
 two years. These two decades are expected to encompass the mass of the dark matter
 axion.
\label{fig:ADMXreach}}
\end{figure}

At present, the largest detector of this type is the Axion Dark Matter
experiment (ADMX), operated at the University of Washington~\cite{Asztalos:2011bm}. 
This detector is now in the commissioning of its upgrade
to a ``definitive search'', a search able to detect dark-matter
PQ axions or eliminate the hypothesis at high confidence. Over the
next decade, this detector will search the mass region
1$\mu$eV--100$\mu$eV even should PQ axions be a minority fraction
of our Milky Way dark-matter halo. The ADMX search in the second mass decade
will require a substantial investment in an upgraded experiment
insert, but not the extensive development of new technology.
Although not theoretically favored to
be dark-matter, the mass decade 100$\mu$eV--1000$\mu$eV is
still unexplored. Such high mass (and therefore high frequency)
axions are the focus of an R\& D effort in developing
suitable high-frequency tunable structures and noise quantum-limited
receivers or bolometric instrumentation. The mass and coupling reach
of ADMX is show in \ref{fig:ADMXreach}

RF cavity axion searches would be difficult to implement at frequencies below a fee 100 MHz
due to the large size of the corresponding structures. However, there is interest in going to
this low frequency or even lower, for instance for axions of the anthropic type.
There have been ideas to exploit the time dependence of the local axion laboratory
field to search for time dependent CP violating interactions. Recently, an experiment
was proposed to look for the spin precession from CP odd nuclear moments via NMR
techniques~\cite{Budker:2013hfa}. This work is in the R\& D phase, and detailed
estimates of noise sources are still being explored. But this idea and its variants
are a new approach to expand the axion search space.

B. Laser Searches I: Shining Light Through Walls.

Axions are pseudoscalar particles, and an electric field
crossed with a magnetic field is likewise pseudoscalar.
Hence, photons of an appropriate polarization traveling through
a transverse magnetic field can convert into axions.
These axions may then leave the beam, thereby depleting one
polarization component, or the axions may reconvert into
photons in a second magnet. Should axions reconvert into
photons within the original magnet, the magnetic field
plus virtual axions introduces a birefringence
to the vacuum. In the ``Shining Light Through Walls'' experiments,
polarized laser light is directed down the bore of a transverse dipole magnet.
The light is then blocked by an opaque wall. Some of the photons
convert into axions, and these axions easily pass through the
wall and reconvert to photons in a second dipole magnet.
The photon-axion-photon conversion rate is very small, since
the axion to two-photon coupling is so tiny, and the entire
photon-axion-photon process contains the product of two
such tiny couplings. Such experiments are unlikely to be sensitive
to PQ type dark-matter axions and are less sensitive than the
SN1987A bound. These experiments are therefore more fully considered
in the Intensity Frontier~\cite{VanBibber:1987rq}.

More recently, experiments are being proposed and are under construction
that increase the conversion rate by introducing a pair of locked Fabry-Perot
optical cavities on either side of the wall. The
conversion rate is thereby enhanced by approximately
the product of the cavity finesses, with the sensitivity
improving as the square-root of this rate~\cite{Mueller:2009wt}. 
A large experiment based on this locked pair of optical
cavities is REAPR, a project proposed for US funding, but not year approved.
A second large experiment ALPS II (proposed for construction in several
phases) has started construction at DESY. These experiments have
improved sensitivity, but are unlikely to reach sensitivity to
PQ type dark-matter axions.

A variant of this idea exploits the absorption and emission
of axions within electromagnetic fields within a single high finesse
Fabre-Perot optical cavity.
Here, a carrier signal is applied to the cavity and the axions
would introduce sidebands on the carrier. The proponents argue
this would be sensitive to an optimistic PQ type dark-matter axion~\cite{Melissinos:2008vn}.

C. Laser Searches II: Dichroism and Birefrigence.

As mentioned in the previous section, photons in light entering a transverse
magnetic field may convert into axions, depleting photons
depleting polarization transverse to the magnetic-field direction, thereby
inducing vacuum dichroism. These axions may then reconvert
into photons within the same magnet. This special direction of
polarization for the conversion and reconversion alters the
propagation velocity of one beam polarization, thereby
introducing vacuum birefringence. Both effects lead to
conversion of linearly polarized into elliptical polarization.
Such rotation of the beam polarization may be detected by
sensitive optical ellipsometers.

Since high-order QED will also mimic such effects, an axion search
can be done with the same apparatus. The prediction for axions
is dichroism is a stronger contributor at lower axion masses,
and birefringence is stronger at heaver masses. This method was briefly
in the spotlight in 2005 when PVLAS reported detecting
vacuum dichroism that could be interpreted as the effect of an
axion in the neighborhood of 1000$\mu$eV with rather large couplings
to photons, couplings that were in tension with the searches mentioned
in the previous section. Further calibrations and studies on the
detector cast doubt on the original claim. The collaboration then rebuilt
the apparatus and failed to reproduce that result~\cite{Zavattini:2007ee}.

Summarizing, this technique is unlikely to have sensitivity to PQ dark-matter axions.
The resulting limits are substantially less sensitive than that from SN1987A, and
there are no plans for a larger-scale experiment.

D. Direct Detection of Solar Axions

As discussed, axions may be produced in the Sun, then propagate
to Earth and when converted to photons in a detector, appear as an
excess of x-rays from the direction of the Sun.
The terrestrial detector consists of a dipole magnet with
bore steered in the direction of the Sun, plus x-ray detectors
at the end of the bore. This is dubbed a ``helioscope''.

These detectors have gone through several generations and are now
highly developed.
The most sensitive such experiment is  the
CERN Axion Solar Telescope (CAST)~\cite{Aalseth:1999ch}
The CAST apparatus consists of an LHC main-ring dipole magnet
on an steerable alt-az mount. The x-ray detection hardware include
grazing-incidence x-ray focusing and ``micromegas'' x-ray detection.
To vary the axion-to-photon dispersion relation, the magnet bore
can be filled with various gasses at various pressures.
This detector recently became sensitive to plausible PQ dark-matter
axions at the upper end of the allowed mass window (around 1~meV).
In general, CAST limit are at or are slightly better than the
red giant bounds.

A larger and more sensitive $4^{\rm th}$ generation helioscope, the International Axion Observatory
(IAXO,~\cite{Vogel:2013bta, IAXOLOI}, has been proposed. Its improved sensitivity is due to a purpose-built
large magnet, X-ray focusing optics and low-background detectors. 
IAXO will be $\approx 20$ more sensitive to the axion-photon coupling constant $g_{a\gamma}$ than CAST, thereby reaching
a few $\times 10^{-12}$~GeV$^{-1}$ regime for a wide range of axion masses up to about $0.25$ eV.
IAXO will therefore be sensitive to a region of the QCD axion parameter space in the third unexplored mass decade
and is sensitive to certain models of non-PQ axion and light-particle dark matter, plus
other weakly-interacting hypothetical particles. The IAXO search
region also explores the axion interpretation of the anomalous white dwarf cooling hints and
anomalously long light propagation over cosmological distances.
This is a large multi-purpose facility and more details on the instrument and
science reach may be found in the IAXO letter of intent.

The history of solar axion searches includes germanium and other
scintillation detectors. Such detectors convert axions to photons
when the Bragg angle within the crystal is satisfied~\cite{Avignone:1997th}
The resultant
limits are considerably less sensitive than that from other
astrophysical bounds. However, such axion searches will likely
continue as an adjunct to germanium WIMP recoil searches.

Finally, one can use data from an x-ray satellite as an on-off
measure when the Earth shields the Sun. Axions emitted by the
Sun would penetrate the Earth and after penetration convert
to photons in the Earth's magnetic field. The resulting sensitivity
would be about that of the red giant bound~\cite{Davoudiasl:2005nh}
A refinement of this is search for x-rays from axions
converting into x-rays in the heliosphere. The x-rays are
then detected by satellites. Preliminary studies have been
done, but it's unclear what the expected sensitivity will be.

E. Short Distance Gravity.

Axions and other light bosons would introduce new short-range
forces. These forces may have a Yukawa form and become comparable
to gravitational forces for distances at 100$\mu$m separations
or less. The force may be between masses, between mass and spin,
or between spin and spin. Axions, being a pseudo scalar, would
dominantly contribute to the mass and spin interaction. A
scalar would have a dominant contribution to the mass-mass
interactions~\cite{Moody:1984ba}
The results from mass-mass couplings are now
severely constrain dark-matter composed of scalars. They also
require dark matter closely obey the equivalence principle
and respect Lorentz invariance. However,
experiments to measure the mass-spin contribution are
less constraining and do not yet approach the required
sensitivity to detect PQ type dark-matter axions. These experiments
are being continuously improved and refined, but there are no proposals
for a significantly expanded program for short-distance gravity
detection of dark-matter~\cite{Hoedl:2011zz}

\noindent\leftline{\underline Astrophysical Bounds}

As mentioned earlier, axions can effect the evolution of or
energy transport in astrophysical objects, or alter the propagation
of photons through intergalactic space. These astrophysical bounds,
especially the neutrino signal from SN1987A and the luminosity
function of white dwarfs are the main observational/experimental
constraint on PQ dark-matter axions: they require such axions to
have masses below about a few times 1000$\mu$eV.

\noindent{A. Energy Transport in Astrophysical Objects}

Axions and other kinds of low-mass particles, very weakly interacting
particles are produced in the hot interiors of astrophysical objects
and thereby become a new channel for energy transport.
This new energy transport channel can, for instance, alter the evolution of
the astrophysical object, so that its evolutionary age
is in conflict with its clock age. This allows inferences to be
made on the properties of these weakly interacting particles, including
their couplings to normal matter and radiation.

\noindent{\sl Energy Transport in the Sun}

KSVZ axions are produced predominantly in the core of the Sun by
photon to axion conversion in scatters off the potential of solar nuclei.
Since the core of the Sun is hot, the axion energy spectrum
peaks in the x-ray. These axions may be detected on Earth directly.
Alternatively, the energy transport of axions from the Sun, combined
the constraint on the solar luminosity and the neutrino fluxes
from SNO provide limits to the axion coupling. The axion emission may
also alter the solar temperature and thereby the solar density
and the solar seismic modes. Approximately, these bounds
are considerably less sensitive to PQ dark-matter axions than
the bound from SN1987A~\cite{Gondolo:2008dd} However, there are non-PQ axion and
other models where these methods have good sensitivity.

\noindent{\sl Energy Transport in Red Giants}

Stars on the red giant branch of the HR diagram eventually
reach the horizontal branch, where helium burning becomes
appreciable and axion emission is also appreciable. The
energy released by axions greatly accelerates the evolutionary
age of the horizontal branch stars, thereby depleting their
population on the horizontal branch. The resulting bound
on the axion coupling is somewhat more restrictive than
that of solar axions (though terrestrial searches for
solar axions have recently bettered this bound). Again,
these bounds are considerably less sensitive to PQ dark-matter
axions than that from SN1987A,though they may have good
sensitivity to certain non-PQ axion and other models.

\noindent{\sl Energy Transport in White Dwarfs}

Nothing forbids the axion to have direct couplings to electrons
or other leptons,
in which case the axion may be emitted as Bremsstrahlung
radiation from hot electrons in astrophysical objects.
This axion to electron coupling is hard to predict, as it
depends directly on the axion's electromagnetic anomaly
which is highly model dependent. The plasma frequency
in white dwarfs is relatively large, which suppresses
the axion Bremsstrahlung channel. The new axion energy
channel would accelerate the white dwarf cooling.
Interestingly, there are suggestions that indeed there
is an extra component to white dwarf cooling, which
could be interpreted as due to a light axion~\cite{Isern:2013lwa}
This warrants more study, as the effect is subtle and the
white dwarf luminosity function is not predicted to high precision.
Similarly, the ZZ-Ceti subclass of variable white dwarfs have a
period that depends on temperature. The period decreases over time,
and the rate of this decrease can be related to energy transport.
Also interestingly, there could be a slight amount of additional
energy loss, which can be interpreted as above as due to
a light axion~\cite{Isern:2010wz}
Also, again, this warrants more study as the
effect is subtle and the period luminosity relation is
not predicted to high precision.

\noindent{\sl Energy Transport in Supernovae}

The most restrictive experimental bound on PQ dark-matter
axions come from the width of the arrival-time burst of
neutrinos from SN1987A. The hot interiors of supernovae
can release an appreciable amount of energy in axions
and other light weakly-interacting particles. This
energy emission rate is can be predicted with reasonable
precision. Axions with too strong a coupling are trapped
and have little observable effects,
while axions with too weak a coupling are rarely produced
and likewise have little effect. Axions whose interaction
length is on order of the size of the supernovae core
release the most energy. The signature of this energy
release is a modification of the neutrino arrival time
burst from SN1987A. Recall, approximately 20 neutrinos were
detected in the approximately 10 second burst. While perhaps
in detail, the arrival time distribution varies from the
predicted arrival time distribution, the overall number of
neutrinos and the overall burst duration closely matches
expectation from a core collapse supernova. The luminosity
in axion emission is thereby constrained, as are then
the axion couplings~\cite{Raffelt:2006cw} Such axions neatly fill the
range between the upper end of the allowed mass window
and other experimental bounds.

Several comments are in order. Firstly, this supernova bound
is very constraining. It rules out PQ axions
with masses above about 1000$\mu$eV. Further, below this mass,
the SN1987A bound is much more sensitive than other astrophysical bounds
and terrestrial experiments (with the exception of RF cavity
experiments and possibly the white dwarf bound discussed earlier).
Secondly, the weak evidence for an extra axion white dwarf
energy transport channel cooling by axion
emission discussed above may not be excluded by the SN1987A
bound.

\noindent{C. Optical and Radio Telescope Searches}

Historically, there was a gap in the otherwise seamless set
of limits for axions with masses above around 1000$\mu$eV.
Such axions, if the dark matter, would make up halos of
galaxies and very slowly decay into two photons. As the
axion mass increases, the axion decay width increases, so that
eV mass axions in halos may be detectable through their glow.
The glow would be an almost monochromatic line (modified by
the Doppler virial velocity). Such emission was not seen,
which closed the gap~\cite{Bershady:1990sw}
It's now appreciated such relatively
heavy axions would be warm dark matter and be highly
constrained.

A variant of this technique is to look for the radio emission from
light axions in halos of dwarf galaxies. The dwarf galaxies
have low virial velocities and hence the emission line
is very narrow on the receiver baseline noise.
A search was performed over a narrow mass range in the
second ``invisible axion'' mass decade~\cite{Blout:2000uc} There is considerable
room for refinement in this method: the collecting area
can be increased, and multi-dish correlation can reduce
the receiver-noise baseline.

\noindent{D. Propagation of Astrophysical Photons}

Intergalactic astrophysical magnetic fields could mediate
the conversion of axions to photons~\cite{Hooper:2007bq}. Here, the conversion
region is large, while the magnetic fields are small.
For mixing lengths to be sensibly long, the axion mass
needs to be quite small, typically much smaller than
the lower bound of the ``invisible axion'' window.

One search method is to look for anomalous dimming
of distant sources. The actual signature could be
quite complex, involving polarization effects and
frequency dependence. At one time it was posited
that the dimming from distant supernovae could be
due to ultra-light axion conversions. This is now
believed to be a negligible effect compared to
the dimming from cosmic acceleration.
It has more recently been suggested there's anomalous
TeV gamma ray transparency from distant AGNs, and
perhaps this is due to conversion of gamma rays to axions,
then back to gamma rays~\cite{DeAngelis:2007dy,Simet:2007sa}.
However, inclusion of secondary gamma rays produced in cosmic-ray
interactions along the line of sight explains the apparent 
transparency without the need for hypothetical new 
particles~\cite{Essey:2009zg,Essey:2009ju,Essey:2010er,Essey:2010nd,Essey:2011wv,Murase:2011cy,Razzaque:2011jc,Prosekin:2012ne,Aharonian:2012fu,Zheng:2013lza,Kalashev:2013vba,Takami:2013gfa,Inoue:2013vpa,Essey:2013kma}. This is again a case where
more study is warranted. This is discussed in the
cosmic ray section of the Cosmic Frontier report.

Along these same lines, it been suggested that axions emitted
from SN1987A would then have reconverted into gamma rays in
intergalactic magnetic field. Hence SN1987A would have been
accompanied by a gamma ray burst. No such burst was seen, 
which provides another bound similar in spirit to the
neutrino SN1987A bound.
There have also been suggestions that similar studies could be
done on other astrophysical objects with high magnetic fields,
but these searches have yet to mature and their potential
sensitivity is unknown.

\noindent{E. Axion Summary}

\begin{figure}[!htb]
  \includegraphics[width=\textwidth]{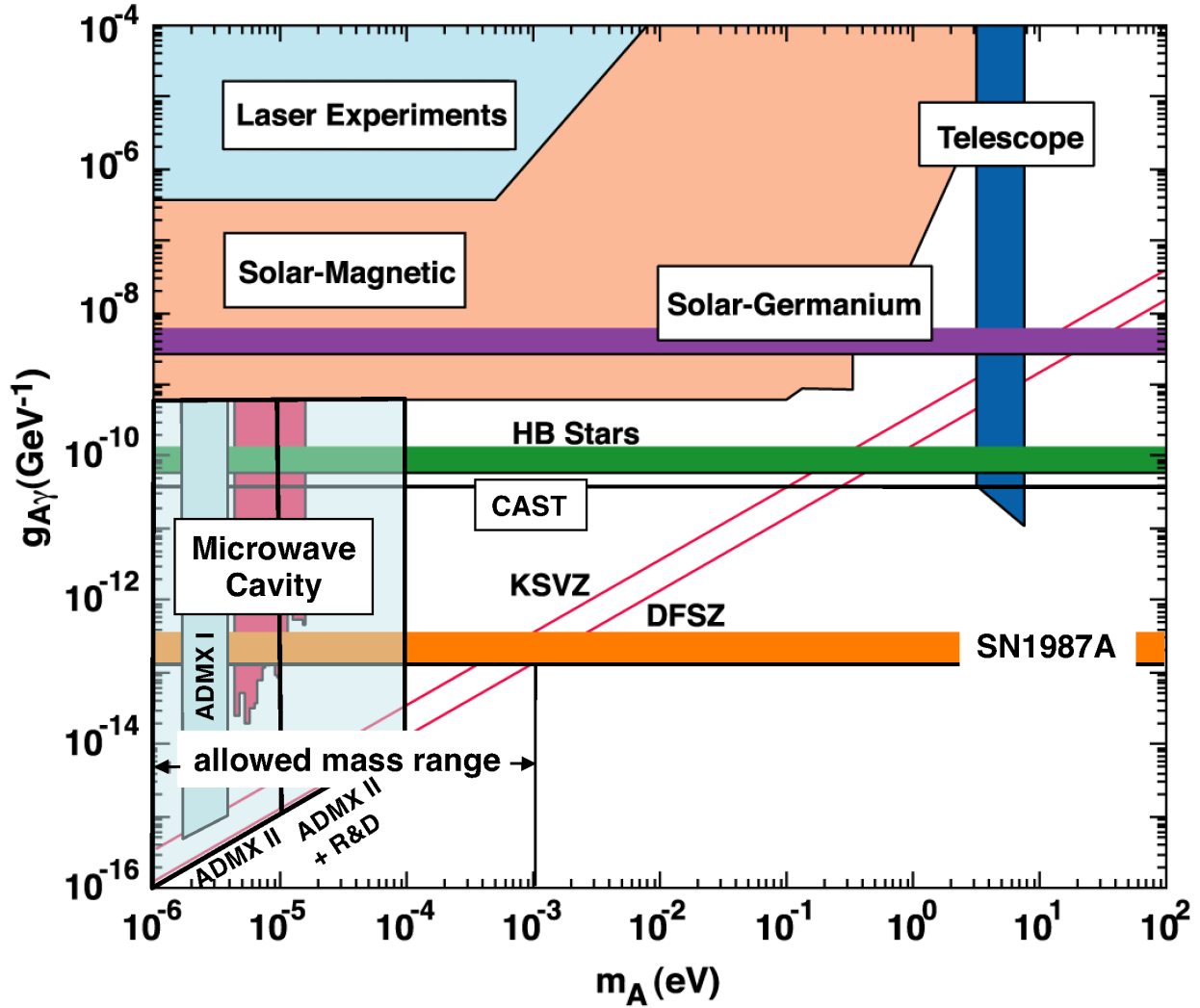}
\caption{The landscape
of axion searches. The vertical axis is the axion's coupling
to two photons. The horizontal axis is the axion's mass.
The diagonal lines are the expected range in coupling for the QCD axion.
The allowed QCD axion window is approximately between 1~$\mu$eV
and 1~meV. Dark matter QCD axions are in the approximate mass range
1~$\mu$eV to 100~$\mu$eV, with the bounds having considerable
uncertainties.
Also shown are upper limits from SN1987A (also white dwarfs) and HB stars
(the red giant bound). Sensitivities of various technologies are
also shown (``Laser'', etc.). The QCD (PQ) dark-matter axions will be explored
with high sensitivity in the next decade by RF-cavity experiments. The solar experiments (CAST and IAXO)
have sensitivity a a large part of the non-PQ search space
and the upper end of the QCD axion window. Of course, there could be surprises
in both mass and couplings.
\label{fig:axionlandscape}}
\end{figure}

The QCD axion has a well bounded parameter
space of mass and couplings. 
There are axion and axion-like-particle alternatives to the QCD axion,
with a vast and largely unexplored search space. The landscape
of this search space is shown in figure~\ref{fig:axionlandscape}.
The diagonal lines are the expected range in coupling for the QCD axion.
The allowed QCD axion window is approximately between 1~$\mu$eV
and 1~meV. Dark matter QCD axions are in the approximate mass range
1~$\mu$eV to 100~$\mu$eV, with the bounds having considerable
uncertainties.
Also shown in the figure are upper limits from SN1987A (also white dwarfs) and HB stars
(the red giant bound). Evolution of stars with masses 8--10~$M_\odot$ provides comparable 
constraints~\cite{Friedland:2012hj}.  Sensitivities of various technologies are
also shown (``Laser'', etc.). The QCD (PQ) dark-matter axions will be explored
with high sensitivity in the next decade by RF-cavity experiments. The solar experiments (CAST and IAXO)
have sensitivity a a large part of the non-PQ search space
and the upper end of the QCD axion window. Of course, there could be surprises
in both mass and couplings.

\subsection{Black holes}  
\def\msun{{M_\odot}}
\def\ten#1{\times 10^{#1}}
\def\aten#1{10^{#1}}

Primordial black holes (PBH) constitute a viable dark matter candidate.  This is probably the only 
possible form of dark matter that is not made up of exotic new elementary particles or forms of matter. 
Black Holes (BH) have been contemplated as Dark Matter candidates since at
least 1970s~\cite{Hawking:1971ei,Khlopov:1985jw,Khlopov:2008qy}.  In fact, one of the 
initial motivations for the scale-free Harrison-Zeldovich 
spectrum of primordial fluctuations was to avoid creating too many PBHs.
PBHs can easily be made during inflation via a ``tilted spectrum''  of fluctuations.  

While BHs naturally form as the end stage of stellar evolution of massive   s, 
astrophysical black holes cannot make up a significant fraction of dark matter. 
Created as final stages of gravitational collapse of some baryonic matter, astrophysical
black holes cannot have a big enough mass fraction to account for dark matter, because 
DM density exceeds the density of ordinary baryonic matter. 
However, if BH's formed  by a different method before the epoch of matter domination, 
they could make up the entirety of the DM.

Primordial black holes (PBH) could form in the Early Universe from overdensities in
the matter/radiation fluid~\cite{Hawking:1974rv,Hawking:1974sw}.  
Whenever the amount of mass inside the horizon exceeds 
the Chandrasekhar mass, one expects BHs to form, and standard cosmological
theory shows that this mass is very near to this limit throughout cosmic
time.  A density enhancement of a few tens of percent at any time during the early universe 
evolution can be sufficient for PBHs to form.  
There have been many suggestions of ways to get such
density enhancements, for example, from bubble collisions,  
collapse of string loops, features in inflaton potentials, 
various types of phase transitions, etc.~\cite{Frampton:2010sw,Carr:2009jm,Khlopov:2008qy}.

Some mechanisms of PBH creation result in a broad spectrum of PBH masses, 
but one appealing scenario involves two consecutive periods of inflation separated by a short 
phase during which coherent oscillations of the inflaton cause preheating and enhancement of 
density perturbation on the scale comparable to the horizon size~\cite{Kawasaki:1997ju,Frampton:2010sw}.  When these density perturbations 
re-enter the horizon after the second inflationary epoch is over, they cause formation of black holes.  
The mass function of black holes produced in this scenario is very narrow: it is practically a delta function determined  
by the size of the horizon between the two stages of inflation.  In addition to producing dark matter, this mechanism is 
capable of generating the primordial seeds for supermassive black holes that exist in the centers of galaxies~\cite{Kawasaki:2012kn}.

We note that if PBHs are created early enough and in appropriate 
mass ranges they can evade big bang nucleosynthesis and 
CMB constraints and make up the entirety of the DM.
PBH lighter than $10^{15}$~g would have decayed by Hawking radiation in less than 10~Gyr. 
Theoretical predictions for the range of PBH masses are not precise, and all masses
in the range from $10^{15}$~g to $10^{38}$~g can be generated from inflation or another mechanism.

While there are many theoretical ideas on how to create PBH DM,
none of them is predictive enough to pinpoint the mass range of the
resulting PBH DM.  Observational constraints eliminate most of the mass range, 
except for a window between $3\ten{-13}\msun$ and $2\ten{-8}\msun$, where the lower limit is set
by recent femto-lensing results from the Fermi-Gamma-ray Burst Monitor data~\cite{Barnacka:2012bm}, and the upper limit is the 
combined MACHO/EROS constraints due to microlensing \cite{Alcock:1998fx}.

To explore the remaining window, several new ideas have been put forward.  
For example, at the upper end of the window between
$3\ten{-13}\msun$ to $2\ten{-8}\msun$ new microlensing experiments
can play an important role.  An analysis of Kepler satellite
data show that this could reach perhaps 40\% of this window,\cite{Griest:2011av}
and recent preliminary results from the same group
seems to eliminate PBH masses down to $2\ten{-9}\msun$.
Using data from upcoming satellite missions should allow further progress 
to be made in exploration of this window~\cite{Cieplak:2012dp}
At the low end of the large PBH mass window, there have been several claims
that the existence of neutron stars in globular clusters eliminates the
possibility of PBH DM in the mass range between the femto-lensing limit
and $2\ten{-9}\msun$~\cite{Capela:2013yf}.  
However, these results have been questioned, 
since the assumed amount of DM in the globular clusters is much larger than
standard astrophysical models suggest.  
One can expect improvement in the femto-lensing
limit due to analysis of FERMI satellite data\cite{Barnacka:2012bm}.  
Accretion onto PBH DM is expected to distort the CMB
and strong limits have been claimed due to non-observation of this 
effect\cite{Ricotti:2007au}. 
However, this analysis was based on strong assumptions, and the limits may 
not be as strong~\cite{Carr:2009jm}.

In summary, PBH remains a viable candidate for dark matter.  PBHs can be created via scalar dynamics in the early Universe.
The remaining mass window can probably be fully explored in the near future.

\subsection{Mirror dark matter} 
\label{sec:mirrorDM}

Mirror dark matter may today be viewed as a special case of asymmetric dark matter, although its original motivation was quite 
different from that of typical asymmetric dark matter models constructed in recent years.  Its origins are also quite old, 
going back to the seminal paper by Lee and Yang on parity violation in the late 1950s~\cite{Lee:1956qn}.  
The basic idea is that despite the $V-A$ character of weak interactions, the true microscopic theory of fundamental interactions might be completely 
symmetric under spatial inversion, $\vec{x} \to -\vec{x}$.  In its purest form, both the 
Lagrangian \emph{and the vacuum} are parity symmetric~\cite{Foot:1991bp}, differentiating mirror matter models from left-right symmetric models.  
While the latter have parity-invariant Lagrangians, experimental constraints force one to spontaneously break that discrete symmetry.  
Mirror dark matter has been argued~\cite{Foot:2003iv,Foot:2010hu,Foot:2011pi,Foot:2012rk,Foot:2013msa} 
to provide a compelling explanation for the positive direct-detection claims of DAMA~\cite{Bernabei:2008yi,Bernabei:2010mq}, 
CoGeNT~\cite{Aalseth:2010vx,Aalseth:2011wp}, CRESST~\cite{Angloher:2011uu} and CDMS/Si~\cite{Agnese:2013rvf}.  
This motivates the consideration being given to this special dark matter candidate.

The definition of a mirror matter model is as follows.  Let ${\cal L}(\psi)$ be a Lagrangian that describes the visible world of standard 
elementary particles $\psi$ through a theory with gauge group $G$.  (The set of fields $\psi$ contain scalar, fermion and gauge fields.) 
In the following, we take this theory to simply be the 
standard model augmented by some neutrino mass-generation mechanism.  Now consider an independent 
but isomorphic gauge group $G'$, with a set of fields $\psi'$ that transform under $G'$ 
in exactly the same way that standard particles $\psi$ transform under $G$.  The $G \times G'$ gauge theory
\begin{equation}
{\cal L}(\psi) + {\cal L}(\psi') + {\cal L}_{\rm int}(\psi, \psi')
\label{eq:mirror-matter-Lag}
\end{equation}
defines a \emph{mirror matter model}, provided that ${\cal L}(\psi')$ is an exact copy of ${\cal L}(\psi)$ so that a non-standard parity 
symmetry $P'$ exists, under which $\vec{x} \to -\vec{x},\ \psi \leftrightarrow \psi'$.  The interaction Lagrangian ${\cal L}_{\rm int}$, 
that depends on both ordinary and mirror fields, is also required to obey this discrete symmetry.  Note that LH (RH) ordinary fermions must 
transform into corresponding RH (LH) mirror fermions under the parity transformation $P'$.  
A non-standard exact time-reversal transformation $T'$ may be defined according to $P' T' \equiv CPT$, 
where the right-hand side is just the standard $CPT$ symmetry obeyed by all local, relativistic quantum field theories.  
A mirror matter model is therefore automatically invariant under all of the improper Lorentz transformations.  
This simple, aesthetically-motivated symmetry requirement is the fundamental motivation for mirror matter.

Among the fields $\psi$ and $\psi'$ are the standard Higgs doublet $\Phi$ and its mirror partner $\Phi'$.  It is easy to show that the minimal scalar 
sector comprising just these two scalar multiplets has only three possible vacuum expectation value patterns, 
depending on the chosen scalar potential parameter region~\cite{Foot:1991bp}: 
\begin{eqnarray}
\langle \Phi \rangle & = & \langle \Phi' \rangle = 0\ ({\rm no\ symmetry\ breaking})\ , \\
\langle \Phi \rangle & \neq & 0,\ \ \langle \Phi' \rangle = 0\ \  {\rm or}\ \ \langle \Phi \rangle = 0,\ \ \langle \Phi' \rangle \neq 0\ ({\rm single\ sector\ symmetry\ breaking})\ , \\
\langle \Phi \rangle & = & \langle \Phi' \rangle \neq 0\ ({\rm parity\ symmetric\ symmetry\ breaking})\ .
\label{eq:Psym-vacuum}
\end{eqnarray}
The vacuum of most interest is given by Eq.~(\ref{eq:Psym-vacuum}) because it leaves the improper spacetime transformations exact while spontaneously breaking the 
electroweak and mirror-electroweak gauge symmetries at the same scale.

No mention of dark matter was made in defining mirror matter models, only the symmetry principle that improper Lorentz transformations are 
to be exact symmetries of nature.  But for free one obtains a hidden sector with exactly the same microphysics as the ordinary or visible sector.  
Given the vacuum of Eq.~(\ref{eq:Psym-vacuum}), all ordinary particles have exactly mass-degenerate mirror partners (note that mass eigenstate 
neutrinos and physical Higgs bosons will in general be admixtures of ordinary and mirror states). The stable particles of the visible sector 
(protons, electrons, some bound neutrons, the lightest neutrino, photons, gluons) have corresponding stable mirror partner 
particles; the exact $P'$ symmetry tells us that the dark matter mass scale is that of the proton and heavier nuclei, 
which is of considerable interest in light of the few-GeV mass scale favored by the positive direct-detection signals observed 
by DAMA, CoGeNT, CRESST and CDMS/Si.  The dark mirror world contains all the complexities and richness of the visible sector: 
atoms, molecules, nuclei, and radiation.  If the dark matter is mirror matter, one also expects some of it to manifest as mirror stars and 
mirror galaxies.

Ordinary and mirror matter will interact via gravity and the non-gravitational interactions described by ${\cal L}_{\rm int}(\psi,\psi')$.  
For the case of the mirror minimal standard model, the interaction Lagrangian contains just 
two gauge-invariant terms: kinetic mixing between the hypercharge gauge boson and its mirror partner,
\begin{equation}
\epsilon F^{\mu \nu} F'_{\mu \nu}\ ,
\label{eq:kinetic-mixing}
\end{equation}
and the scalar potential term $\lambda_{\Phi\Phi'} \Phi^{\dagger} \Phi \Phi'^{\dagger} \Phi'$, where $\epsilon$ and $\lambda_{\Phi\Phi'}$ are arbitrary dimensionless 
parameters.  If gauge-singlet neutrino-like states exist, they will in general mix and constitute a third non-gravitational connection 
between the sectors.  The limit $\epsilon, \lambda_{\Phi\Phi'} \to 0$ is technically natural in the absence of gravity because the sectors decouple, 
with independent Poincar\'{e} transformations as the enhanced symmetry.  This fact justifies very weak 
non-gravitational couplings between the ordinary and mirror sectors, as required for mirror matter to be dark matter.  Nonetheless, 
a small but nonzero value for $\epsilon$ is important for the testability of this hypothesis, and perhaps even for its cosmological viability.

Indeed, does mirror matter succeed phenomenologically as a dark matter candidate?  The answer is ``yes'', at least within the uncertainties 
caused by the great complexity of mirror dark matter and the associated calculational challenges.  To see why this answer is reasonable, 
one has to understand mirror dark matter cosmology~\cite{Berezhiani:2000gw,Ignatiev:2003js,Berezhiani:2003wj,Ciarcelluti:2004ik,Ciarcelluti:2004ip,Ciarcelluti:2008qk,Foot:2012ai}.

The fundamental fact to be appreciated about this cosmology is that the ordinary and mirror plasmas of the early universe should have 
thermally decoupled prior to big bang or primordial nucleosynthesis and have different temperatures $T$ and $T'$, respectively.  
Successful primordial nucleosynthesis requires that the mirror plasma temperature be slightly smaller than the temperature of the 
ordinary-particle bath, $T'/T \stackrel{<}{\sim} 0.5$, in order to meet the upper bound on extra radiation during that epoch~\cite{Berezhiani:2000gw,Ignatiev:2003js}.  
This phenomenologically-necessary temperature difference, possibly set up by an inflationary mechanism~\cite{Berezhiani:2000gw}, 
makes the evolution of complexity and structure in the mirror world quite different from that of standard matter, and explains why mirror 
dark matter manifests in an observationally different way from ordinary matter in cosmology and astrophysics despite the identical microphysics. The disparities begin with primordial nucleosynthesis, where the lower temperature $T'$ means that the ratio of mirror-Helium He$'$ to mirror-Hydrogen H$'$
is much higher than the corresponding ordinary-sector figure.

Can mirror dark matter lead to successful large-scale structure formation?  As it happens, the lower temperature is an important consideration.  
To avoid a delay in the onset of structure formation, the mirror photons should decouple from the mirror nuclei and electrons no later than the time of 
matter-radiation equality, which in turn motivates that $T'/T \stackrel{<}{\sim} 0.3$, 
a little lower than the maximum permitted by primordial nucleosynthesis~\cite{Berezhiani:2000gw,Ignatiev:2003js}.  
In the linear regime of density perturbation growth, mirror dark matter 
behaves more closely like cold dark matter as the temperature ratio is taken smaller, with values below $0.3$ favored.  
Eventually, of course, the growth of mirror dark matter structure deviates strongly from that of standard, non-dissipative, 
collisionless cold dark matter, with the formation of mirror stars and other condensed structures.

A challenge for mirror dark matter is the need for dark matter halos around spiral galaxies such as the Milky Way to be spheroidal.  
One could ask why the mirror matter should be distributed in this way while the microphysically-identical ordinary matter has collapsed into a disk.  
An answer may involve a heating source to sustain a pressure-supported spheroidal halo.  As it happens, 
ordinary supernovae have suitable energetics to play this role~\cite{Foot:2004wz}, provided the kinetic-mixing 
parameter $\epsilon$ is about $10^{-9}$, a figure that, interestingly, is also motivated by the mirror dark matter 
explanation of the positive direct-detection claims (see below).  Through kinetic mixing of photons and mirror photons, 
about half of the gravitational binding energy released during an ordinary core-collapse supernova can be converted 
into mirror particles, which heat the ionized mirror matter making up the galactic halo sufficiently to solve the problem.  
The reverse process plausibly does not happen, at least in the current epoch, because the plasma is far too hot for significant mirror star formation to occur.

There are constraints on dark matter self-interactions from elliptical galaxies\cite{Feng:2009mn} and the bullet cluster system~\cite{Markevitch:2003at}.
Considering first elliptical galaxies, observations indicate that the dark matter is flattened, i.e.\ non-spherically distributed (see, for example, Ref.~\cite{Buote:2002wd}). This ellipticity constrains self-interacting, nondissipative dark matter, which tends to form spherical halos.
Dissipative dark matter, on the other hand, can collapse to a flattened disk and bulge if there is little supernova heating (expected 
since elliptical galaxies are observed to be devoid of gas and have very low star formation rate).
The Bullet cluster potentially constrains the self interactions of mirror dark matter. However, given the weak constraints
on the proportion of mirror dark matter in the form of intergalactic gas and also
the fairly low self-interaction cross-section at the estimated mirror plasma temperature ($T' \sim 10$ keV), mirror dark matter remains viable.
See Refs.~\cite{Silagadze:2008fa,Foot:2013vna} for further discussions. 

If kinetic mixing exists, with $\epsilon \sim 10^{-9}$, then mirror particles from the halo of the Milky Way can Rutherford 
scatter off of ordinary nuclei on Earth. This leads to the possibility that these particles can be probed in direct detection experiments. 
Although the dominant He$'$, H$'$ halo components are too light to be important for the currently operating
experiments, heavier metal components such as O$'$ or Fe$'$ can be detectable. To determine the interaction rate 
of such a heavy metal component, A$'$, in experiments such as DAMA and CoGeNT, 
one needs to know the velocity distribution of A$'$ and its scattering cross section with ordinary matter.
The mirror particles are presumed to form a pressure supported halo with common temperature $T$.
The halo temperature is set by the Galactic rotational
velocity, $v_{\rm rot} \sim 220$ km/s, via $T \approx {1 \over 2} \bar m v_{\rm rot}^2$, where $\bar m$ is the mean mass of the 
halo mirror particles~\cite{Foot:2004wz}. A novel  feature of such multi-component dark matter is a mass dependent
velocity dispersion, $f({\rm A}') = \exp(-E/T) = \exp(-v^2/v_0^2)$ where $v_0 = \sqrt{2T/m_{{\rm A}'}} = 
\sqrt{\bar m/m_{{\rm A}'
}}\ v_{\rm rot}$. The important point is that heavy mirror particles are expected to have quite narrow velocity dispersion: $v_0^2 \ll v_{rot}^2$. This can be contrasted with WIMPs, which have $v_0 = v_{\rm rot}$ in the standard halo model.  Another distinguishing feature of mirror particles is that they are expected to interact via kinetic-mixing-induced Rutherford scatting.  
The differential cross section for a mirror nucleus, A$'$, with
atomic number $Z'$ and velocity $v$
to elastically scatter off an ordinary nucleus, A, with
atomic number $Z$ is:
\begin{eqnarray}
{d\sigma \over dE_R} = {2\pi \epsilon^2 Z^2 Z'^2 \alpha^2 F^2_{\rm A} F^2_{{\rm A}'} \over m_{\rm A} E_R^2 v^2}
\label{cs}
\end{eqnarray}
where
$F_{\rm A}$ ($F_{{\rm A}'}$) is the form factor
of the nucleus (mirror nucleus) and natural units are used.
Here $E_R$ is the recoil energy of the target nucleus, A.
 
The data from the DAMA, CoGeNT, CRESST-II and CDMS/Si experiments
have been analyzed within this mirror dark matter framework~\cite{Foot:2003iv,Foot:2010hu,Foot:2011pi,Foot:2012rk,Foot:2013msa}.
It has been found that the positive signal from each of these
experiments can be explained with ${\rm A}' = {\rm Fe}'$ and $\epsilon \sqrt{\xi_{{\rm Fe}'}} \approx 2 \times 10^{-10}$ (where $\xi_{{\rm Fe}'}$ is the mass fraction of Fe$'$ in the halo). 
Other regions of parameter space are also possible.
In summary, mirror dark matter will be explored as part of the next generation
WIMP searches and the dark-matter astrophysics program.

\subsection{Self-interacting non-WIMP dark matter \label{sec:self-interDM}}

Dark matter with stronger-than-weak-scale self-interactions arises in a variety of models, which possess their individual motivations, \eg~in mirror DM and more generally asymmetric DM models. However, self-interacting DM is independently motivated as a potential solution to the discrepancies which currently appear between the predictions of collisionless CDM simulations and observations of the galactic and subgalactic structure of the universe, reviewed in Sec.~\ref{sec:astro}.  For the disagreement between simulations and observations to be resolved within the CDM paradigm, some mechanism that modifies the standard CDM structure-formation picture is needed.

If DM interacts significantly inside halos, then the energy transfer among DM particles heats up the low-entropy material concentrated in the center of the galaxies, thus reducing their central densities~\cite{Spergel:1999mh}. Core-type inner density profiles are in better agreement with observations of dwarf galaxies than the cusp-type profiles predicted in collisionless CDM simulations.  Moreover, the reduced central densities imply reduced velocity dispersions, which can alleviate the
``too big to fail problem''~\cite{BoylanKolchin:2011de}.
However, DM self-scattering and energy transfer also tends to form isotropic DM halos. Preserving the observed triaxial nature of elliptical halos gives the most severe constraints on self-interacting DM~\cite{Feng:2009mn,Feng:2009hw}; these constraints are typically stronger than those arising from the Bullet Cluster~\cite{Markevitch:2003at} and from elliptical galaxy clusters~\cite{MiraldaEscude:2000qt}. 
In fact, the galactic and subgalactic clustering patterns of DM appear to be a sensitive probe not only of the strength but also of the nature of the DM self-interactions. 

Dark-matter self-scattering may occur either via short-range interactions with velocity-independent scattering cross sections, or via long-range interactions\footnote{Note that ``long range'' does not imply astronomical distances.  For the case of a massless vector mediator -- a dark photon -- Debye screening due to the DM plasma makes the effective range of the interaction $\lambda_D \approx m_{_\text{DM}} v_{_\text{DM}}/\sqrt{4\pi \alpha_{_D} \rho_{_\text{DM}}} \sim (1 - 10^5) \text{ cm}$ depending on the parameter choice.  A scalar mediator of any reasonable nonzero mass gives rise to a much shorter range, even if that range is large compared to typical particle physics distance scales.} 
with the scattering cross-section decreasing with increasing velocity. 
The momentum-transfer cross-section, defined as  $\sigma_{_T} \equiv \int d\Omega \, (d\sigma_{\chi\chi}/d\Omega) \, (1-\cos \theta) $ with $d\sigma_{\chi\chi}/d\Omega$ being the differential DM self-scattering cross-section and $\theta$ the scattering angle, factors out forward scattering (which does not redistribute energy among DM particles), and is used to parametrize the effect of DM self-interactions in halos.
Recent simulations show that for velocity-independent DM self-scattering cross-sections, DM self-interactions can affect the kinematics of halos without spoiling their triaxiality for a narrow range of values around
$\sigma_{_T} / m_{_\text{DM}} \approx 0.6 \text{ cm}^2/ {\rm g}$~\cite{Rocha:2012jg,Peter:2012jh,Vogelsberger:2012sa,Vogelsberger:2012ku,Zavala:2012us}.
A much broader parameter space is available if the DM self-interaction is long-range.
Because in this case the DM self-scattering becomes suppressed with increasing velocity, its effect is more pronounced in smaller halos with low velocity dispersion, such as the dwarf galaxies, while it becomes unimportant in larger galaxies and clusters, which have much higher velocity dispersions.
References~\cite{Vogelsberger:2012sa,Vogelsberger:2012ku} performed simulations for velocity-dependent cross-sections arising in Yukawa interactions via a light mediator~\cite{Feng:2009hw,Loeb:2010gj}. For benchmark scenarios with roughly $\sigma_{_T} / m_{_\text{DM}} \sim (1-40) \text{ cm}^2 / {\rm g}$ at $v \sim  10-30 \text{ km/s}$, they found that the inner profiles of the subhaloes turned out to be no more dense that what inferred from the kinematics of the dwarf spheroidal galaxies, while the ellipticity of the main halo was retained. Indeed, for the models considered, $\sigma_{_T} / m_{_\text{DM}} \lesssim 0.2 \text{ cm}^2 / {\rm g}$ at velocities $v \gtrsim 100 \text{ km/s}$ relevant to Milky Way and larger size halos, which is consistent with the bound on the transfer cross section in the velocity-independent case.
To delineate the full range of possibilities, more simulations for a wider range of parameters and various types of velocity-dependence of the self-interaction cross section are of course needed.

Several scenarios of self-interacting DM have been explored, including
single component DM interacting via a massless~\cite{Feng:2009mn} or massive~\cite{Lin:2011gj,Tulin:2012wi,Tulin:2013teo} vector boson, or a scalar mediator~\cite{Loeb:2010gj,Tulin:2013teo}, and asymmetric DM coupled to a massless (or very light) vector boson, which gives rise to the atomic DM scenario~\cite{CyrRacine:2012fz}.
Another compelling possibility arises if DM is in the form of $Q$-balls. $Q$-balls can coalesce after a collision, forming larger $Q$-balls and decreasing their number density. This implies that the effective self-interaction rate is reduced to a negligible value after a few collisions per particle~\cite{Kusenko:2001vu}.
Below we describe in some more detail phenomenological aspects of atomic DM and DM interacting via a massive scalar boson.

\paragraph{Atomic dark matter.} 
Asymmetric DM coupled to a massless or light vector boson of a gauged $U(1)_{_D}$ symmetry, is made up of (at least) two species of particles, so that the net gauge charge carried by one species (due to its asymmetric relic abundance) is compensated by an opposite gauge charge carried by the other species.\footnote{
Dark baryogenesis has to occur before the dark baryons --not yet bearing an asymmetry-- have annihilated below the observed DM abundance, i.e. at temperatures 
$T_{_\text{dark BG}} \gtrsim m_{_\text{DM}} / x_f$, where $m_{_\text{DM}}$ is the dark baryon mass, and typically $x_f \sim 20-50$. If the dark gauge symmetry $U(1)_{_D}$ is unbroken, or broken at a scale $M_{_D} \lesssim m_{_\text{DM}}/x_f$, dark baryogenesis occurs before the possible transition of the universe to the broken phase, and by gauge invariance ADM has to consist of both positively and negatively charged particles.} 
This gives rise to the atomic DM scenario, which appears in many ADM constructions,~\eg~\cite{Kaplan:2009de,Kaplan:2011yj,Petraki:2011mv,vonHarling:2012yn,Cline:2012is} (c.f. Sec.~\ref{sec:ADM}).
In analogy to ordinary matter, the two species making up the dark atoms are referred to as the dark proton $p_{_D}$ and the dark electron $e_{_D}$, and the $U(1)_{_D}$ neutral bound state is referred to as the dark Hydrogen $H_{_D}$.
The cosmology of atomic DM consisting of two fermionic species bound by an unbroken $U(1)_{_D}$, has been explored in Ref.~\cite{CyrRacine:2012fz}. The rather rich phenomenology of atomic DM depends on the efficiency of the dark recombination in the early universe
\begin{equation}
p_{_D} + e_{_D} \leftrightarrow  H_{_D} + \gamma_{_D}
\ ,
\label{eq:recombination}
\end{equation}
where $\gamma_{_D}$ is the dark photon, the late-time ionization fraction of DM, the thermal decoupling of DM from the dark radiation, and the evolution of the DM density perturbations.
Even richer phenomenology emerges if the dark sector features also a strong force which binds dark particles into heavier states and gives rise to nuclear physics, as is the case in mirror DM models (c.f. Sec.~\ref{sec:mirrorDM}). 

The cosmology of the atomic DM scenario is determined by four parameters: the dark fine-structure constant $\alpha_{_D}$, the mass of the dark Hydrogen $m_{_D}$, the dark proton-electron reduced mass $\mu_{_D}$, and the present-day ratio of the dark and the visible sector temperatures, $\xi~\equiv~T_{_D} / T_{_V}$. The binding energy of the dark Hydrogen atom is $E_{_D} \simeq \alpha_{_D}^2 \mu_{_D}/2$, and the mass of the bound state is related to the mass of the dark fermions by $m_{_D} = m(p_{_D}) + m(e_{_D}) - E_{_D}$. 
A reasonable estimate for the residual ionization fraction based on equilibrium thermodynamics is
\begin{equation}
x_{\rm ion} \sim  10^{-6} \, \xi
\left( \frac{ m_{_D} \mu_{_D}}{\text{GeV}^2 } \right)
\left( \frac{0.1}{\alpha_{_D}} \right)^4
\  ,
\label{eq:x_ion}
\end{equation}
although the exact value depends on the details of the dark recombination~\cite{CyrRacine:2012fz}.
The residual ionization fraction determines the DM self-interaction inside halos, which involves atom-atom, atom-ion and ion-ion collisions. The corresponding momentum-transfer cross-sections are 
\begin{equation}
\sigma_{_T} \left( H_{_D} \!-\! H_{_D} \right)  \approx 
\frac{120 \pi}{\alpha_{_D}^2 \mu_{_D}^2 v^{1/4}} 
\ , \quad
\sigma_{_T} \left( H_{_D} \!- i \right)  \approx  
\frac{240 \pi} { \alpha_{_D}^2 \mu_{_D}^2 } \left( \frac{m_i}{m_{_D}} \right)^\frac{1}{2}
, \quad
\sigma_{_T} (i-j)  \approx  
\frac{2 \pi \alpha_{_D}^2} {  \mu_{ij}^2 v^4} 
\  , 
\label{eq:sigma atomic DM}
\end{equation}
where $m_i = m(p_{_D}) \text{ or } m(e_{_D})$ are the masses of the dark ions, and $\mu_{ij}$ is the reduced mass of the $i-j$ ion pair. (More accurate expressions for the scattering cross sections are given in Ref.~\cite{CyrRacine:2012fz}.)

Depending on the strength of the dark force, there are various regimes with different phenomenology~\cite{CyrRacine:2012fz}:
\begin{enumerate}[(i)]
\item 
For large values of the dark fine-structure constant, $\alpha_{_D} \gtrsim 0.1$, the cosmology of atomic DM resembles the collisionless CDM scenario. Dark recombination occurs mostly while in thermodynamic equilibrium
and does not depend on the details of the atomic transitions. 
Because the binding energy of the dark atoms is large, dark recombination and the kinetic decoupling of DM occur early, and the matter power spectrum differs from that of collisionless CDM only at unobservably small comoving scales.
The residual ionization fraction is small and the DM self-scattering inside halos is dominated by atom-atom collisions, whose cross-section is insensitive to $v$. Requiring $\sigma/m_{_\text{DM}} \lesssim 1 \text{ cm}^2 / {\rm g}$ to preserve the observed ellipticity of halos implies
\begin{equation}
\alpha_{_D} \gtrsim  0.3 
\left( \frac{10\text{ GeV}}{m_{_D}} \right)^{1/2} 
\left( \frac{\text{GeV}}{\mu_{_D}} \right)
\  ,
\label{eq:alphaD strong}
\end{equation}
where parameter values which are close to satisfying the equality in Eq.~\eqref{eq:alphaD strong} could potentially  resolve the small-scale structure problems of collisionless CDM. 
Because of the large value of $\alpha_{_D}$, the atomic energy splittings are large and the collisions of dark atoms in the halos are not energetic enough to excite them, ensuring that DM is non-dissipative.

\item
For intermediate values of the fine structure constant, the recombination process is in quasi-equilibrium and depends on the details of the atomic transitions. 
Dark acoustic oscillations can imprint a new scale in the matter power spectrum, which determines the minimum DM protohalo mass. A significant residual ionization fraction may be present today, given roughly by Eq.~\eqref{eq:x_ion}. Rutherford scattering of the ionized component, with the strong velocity-dependence of the cross-section, $\sigma \propto v^{-4}$, can potentially alter the halo kinematics, resulting in subhalos with reduced central density, without affecting the ellipticity of the main halo.

\item For very small $\alpha_{_D}$ and/or large dark proton and dark electron masses, the recombination rate is lower than the Hubble rate. The dark sector remains mostly ionized. In fact, dark atoms do not form if
\begin{equation}
\alpha_{_D} \lesssim  10^{-4} \, \xi
\left(\frac{m_{_D}}{\text{GeV}} \right)^{1/4}  
\left(\frac{\mu_{_D}}{\text{keV}} \right)^{1/4}  
\  .
\label{eq:no atoms}
\end{equation}
(However, efficient annihilation in the early universe requires $\alpha_{_D} \gtrsim 3 \times 10^{-5}(m_{_D}/\text{GeV})$.)

\end{enumerate}

\paragraph{Yukawa interactions.}
If DM couples to a light scalar, then its self-interaction cross-section exhibits a non-trivial velocity dependence. For fermionic DM $\chi$ that couples to a scalar $\phi$ via
\begin{equation} 
{\cal \delta L} = g_\chi \phi \bar{\chi} \chi  \ ,
\label{eq:phi chi chi}
\end{equation}
the effect of the DM self-interaction inside halos depends on the three model parameters, the coupling $\alpha_\chi\equiv g_\chi^2/4\pi$, the DM mass $m_\chi$, the mass of the scalar mediator $m_\phi$, and on the velocity $v$ of DM in the halos.
The Born approximation describes adequately the perturbative regime, with the momentum-transfer cross section being~\cite{Feng:2009hw}
\begin{equation}
\alpha_\chi \ll  m_\phi/m_\chi \ : \qquad
\sigma_{_T} = 
\frac{8 \pi \alpha_\chi^2}{m_\chi^2 v^4}
\left[
\ln \left(1+\frac{m_\chi^2 v^2}{m_\phi^2} \right) 
- \frac{m_\chi^2 v^2/m_\phi^2}{1+m_\chi^2 v^2/m_\phi^2}
\right]
\  .
\label{eq:Born}
\end{equation}
At larger couplings, non-perturbative effects become important. In the classical regime, $v \gg m_\phi/m_\chi $, the momentum-transfer cross-section is~\cite{Feng:2009hw}
\begin{equation}
\frac{m_\phi}{m_\chi} \lesssim \alpha_\chi, \
\frac{m_\phi}{m_\chi} \ll v  \ : \qquad
\sigma_{_T} = \left \{
\begin{alignedat}{5}
&\frac{4 \pi \beta^2}{m_\phi^2} \ln \left(1+ \beta^{-1} \right)         \ ,   &\quad & \beta \lesssim 10^{-1} & \\
&\frac{8 \pi \beta^2}{m_\phi^2 \left(1+ 1.5 \beta^{1.65} \right)}     \ ,  &\quad & 10^{-1} \lesssim \beta \lesssim 10^3 & \\
&\frac{\pi}{m_\phi^2} \left( \ln \beta + 1 - \frac{1}{2} \ln^{-1} \beta \right)  \ ,  &\quad & \beta \gtrsim 10^3 & \\
\end{alignedat}
\right.
\label{eq:classical}
\end{equation}
where $\beta \equiv 2 \alpha_\chi m_\phi / m_\chi v^2$. 
Scalar bosons mediate attractive interactions, and for large enough couplings beyond the perturbative regime,  $\alpha_\chi \gtrsim m_\phi / m_\chi$, bound states exist. Outside the classical regime, for $v \ll m_\phi/m_\chi$, bound states can also form and the DM self-scattering exhibits resonances. 
In the resonant regime, the momentum-transfer cross section can be approximated by~\cite{Tulin:2013teo}
\begin{equation}
v \ll \frac{m_\phi}{m_\chi} \lesssim \alpha_\chi  \ : \quad
\sigma_{_T} = \frac{16 \pi}{m_\chi^2 v^2} \sin^2 \delta \  ,
\label{eq:reson}
\end{equation}
where
\begin{equation}
\delta = \arg \left( \frac{i \Gamma \left( \frac{i m_\chi v}{\kappa m_\phi} \right) }{\Gamma (\lambda_+) \Gamma(\lambda_-)} \right) 
\ , \qquad 
\lambda_{\pm} = 1+ \frac{i m_\chi v}{2 \kappa m_\phi} \pm \sqrt{\frac{\alpha_\chi m_\chi}{\kappa m_\phi} - \frac{m_\chi^2 v^2}{4 \kappa^2 m_\phi^2}} \ , \quad \kappa \approx 1.6 \ .
\label{eq:reson param}
\end{equation}

From the above, it follows that for $m_\chi \gtrsim 300 \text{ GeV}$ and $m_\phi \lesssim 30 \text{ MeV}$, the DM self-scattering cross-section can comfortably lie in the range required to affect the kinematics of dwarf galaxies, and also exhibit strong velocity dependence which ensures negligible effect on Milky-Way and galaxy-cluster
scales~\cite{Tulin:2013teo}. This regime is described by the classical approximation of Eq.~\eqref{eq:classical}. A more limited parameter space for
velocity-dependent self-interaction is available in the resonant regime, for $m_\chi \sim (60 - 300) \text{ GeV}$~\cite{Tulin:2013teo}. For lower DM masses, the Born approximation becomes applicable and the self-scattering cross-sections is mostly velocity-independent~\cite{Tulin:2013teo}.

In summary, atomic dark matter has complex self-interactions, as well as interactions with ordinary matter.  Its astrophysical ramifications deserve further studies.  Both direct and indirect detection

\subsection{Sterile neutrinos}
 
Sterile, or right-handed neutrinos are usually introduced to explain the observed neutrino masses; the corresponding new particles can be dark matter in some range of parameters~\cite{Kusenko:2009up}.

The Standard Model was originally formulated with 
massless neutrinos $\nu_\alpha$ transforming as components of the
electroweak SU(2) doublets $L_\alpha$ ($\alpha =e,\mu,\tau$).  To accommodate the
neutrino masses, one can add several electroweak
singlets $ N_{a}$ ($a=1,...,n$) to build a
seesaw
Lagrangian~\cite{Minkowski:1977sc,Glashow:1979nm,Yanagida:1979as,GellMann:1980vs,Mohapatra:1979ia}:
\newcommand{\slashed}[1]{{#1}\hspace{-2mm}/}
\beq  {\mathcal L} = {\mathcal L_{\rm SM}} + i \bar
N_a \slashed{\partial} N_a - y_{\alpha a} H^{\dag} \,  \bar L_\alpha
N_a - \frac{M_a}{2} \; \bar N_a^c N_a + h.c.
\eeq
Here ${\mathcal L_{\rm SM}} $ is the Standard Model Lagrangian (with only the left-handed neutrinos and without the neutrino masses).  We will assume that SU(3)-triplet Higgs bosons~\cite{Schechter:1980gr} are not involved, and all the neutrino masses arise from the ``seesaw'' Lagrangian.

The neutrino mass eigenstates $\nu^{\rm (m)}_i$ ($i=1,...,n+3$) are linear
combinations of the weak eigenstates $\{\nu_\alpha, N_a \}$.  They are obtained
by diagonalizing the $(n+3)\times (n+3)$ mass matrix:
\beq 
{\mathcal M^{(n+3)}} = \left( \begin{array}{cc}
          0 & y_{\alpha a} \langle H \rangle \\
y_{a \alpha} \langle H \rangle & {\rm diag}\{M_1,...,M_n\}
         \end{array}
\right).
\eeq
As long as all $y_{a \alpha} \langle H \rangle \sim y \langle H \rangle \ll M_a\sim M$, the eigenvalues of
this matrix split into two groups:  the lighter states with masses 
\begin{equation}
m(\nu^{\rm (m)}_{1,2,3} )  \sim  \frac{y ^2 \langle H \rangle ^2}{M}  
\end{equation}
and the heavier eigenstates with masses of the order of $M$: 
\begin{equation}
m(\nu^{\rm (m)}_{a} )  \sim  M \ \ \ (a>3).
\end{equation}
We call the former {\em active 
neutrinos} and the latter {\em sterile neutrinos}.  Generically, the mixing angles in this
case are of the order of 
\begin{equation}
\theta_{a\alpha}^2 \sim \frac{y_{a \alpha} ^2 \langle H
\rangle ^2}{M^2},
\end{equation}
but some additional symmetries or accidental cancellations can make them different from these generic values.

One can consider a broad range of values for the number $n$ of sterile neutrinos.  Unlike the 
other fermions, the singlets are not subject to any constraint based on the anomaly
cancellation because these fermions do not couple to the gauge fields.  To explain
the neutrino masses inferred from the atmospheric and solar neutrino experiments, $n=2$ singlets
are sufficient~\cite{Frampton:2002qc}, but a greater number is required if the seesaw 
Lagrangian is to explain the r-process nucleosynthesis~\cite{McLaughlin:1999pd}, the pulsar
kicks~\cite{Kusenko:1997sp,Kusenko:1998bk,Fuller:2003gy,Barkovich:2004jp,Kusenko:2004mm,Loveridge:2003fy,Kusenko:2006rh,Kusenko:2008gh} and the strength of the supernova explosion~\cite{Fryer:2005sz,Hidaka:2006sg}, as well as dark  matter~\cite{Dodelson:1993je,Shi:1998km,Abazajian:2001nj,Abazajian:2001vt,Abazajian:2002yz,Dolgov:2000ew,Asaka:2005an,Kishimoto:2006zk,Asaka:2006ek,Asaka:2006rw}.   A model often referred to as  $\nu$MSM, for Minimal Standard Model (MSM) with neutrino ($\nu$) masses is the above model with $n=3$ sterile neutrinos, all of which have masses below the electroweak scale: one has a mass of the order of a few  keV, while the two remaining sterile neutrino are assumed to be closely degenerate at about 1-10~GeV scale~\cite{Asaka:2005an,Boyarsky:2009ix}.  This model is singled out for the minimal particle content consistent with baryogenesis~\cite{Akhmedov:1998qx,Asaka:2005pn} and having a dark-matter candidate.  However, as discussed below, the need for a cosmological  mechanism capable 
of producing colder dark matter than that generated by neutrino oscillations may require one to go beyond the minimal model~\cite{Merle:2013gea}, and introduce some additional physics at the electroweak scale~\cite{Kusenko:2006rh,Petraki:2007gq} or at a higher 
scale~\cite{Kusenko:2010ik}.

The scale of the right-handed Majorana masses $M_{a}$ is unknown; it can be
much greater than the electroweak scale~\cite{Minkowski:1977sc,Glashow:1979nm,Yanagida:1979as,GellMann:1980vs,Mohapatra:1979ia}, or it
may be as low as a few eV~\cite{deGouvea:2005er}.  Theoretical arguments have been put forth for various ranges of these Majorana masses.  
Some models, including the ``split seesaw'' allow for small Majorana masses~\cite{Kusenko:2010ik}.

The singlet fermions are introduced to explain the observed neutrino masses, but the new particles can make up the dark matter.  Because of the small Yukawa couplings, the keV sterile neutrinos are out of equilibrium at high temperatures.  They are not produced in the freeze-out from equilibrium.  However, there are several ways in which the relic population of sterile neutrinos could have been produced.  

\begin{itemize}
 \item Sterile neutrinos could be produced from neutrino oscillations, as 
proposed by Dodelson and Widrow (DW)~\cite{Dodelson:1993je}.  If there is no lepton asymmetry is
negligible, this scenario appears to be in conflict with a combination of the X-ray
bounds~\cite{Abazajian:2001vt,Abazajian:2005gj,Beacom:2005qv,Mapelli:2005hq,Boyarsky:2005us,Abazajian:2006yn,Boyarsky:2006fg,Boyarsky:2006jm,Watson:2006qb,Abazajian:2006jc,Boyarsky:2006ag,Boyarsky:2006hr,RiemerSorensen:2006fh,RiemerSorensen:2006pi,Boyarsky:2006kc,Boyarsky:2006jm,Yuksel:2007xh,Boyarsky:2007ay,Loewenstein:2008yi,RiemerSorensen:2009jp,Loewenstein:2009cm,Watson:2011dw,Loewenstein:2012px} on one hand,  and the Lyman-$\alpha$ and structure formation bounds on the other hand~\cite{Narayanan:2000tp,Viel:2005qj,Seljak:2006qw,Viel:2007mv,Markovic:2010te,Kamada:2013sh}.  The latter allow no more than a half of dark matter abundance to be in the form of warm sterile neutrinos.\footnote{On the other hand, observations of dwarf spheroidal galaxies point to core-type profiles, which could be a manifestation of a non-trivial velocity distribution of self-interaction of dark
matter~\cite{Kauffmann:1993gv,Hernandez:1998hf,SommerLarsen:1999jx,Klypin:1999uc,Moore:1999nt,Bode:2000gq,Dalcanton:2000hn,Peebles:2001nv,Zentner:2002xt,Simon:2003xu,Gentile:2004tb,Goerdt:2006rw,Strigari:2006ue,Gilmore:2006iy,Wilkinson:2006qq,Boyanovsky:2007zz,Gilmore:2007fy,Wyse:2007zw,Gilmore:2008yp,Koch:2008dc,Munari:2008hb,Siebert:2008uu,Veltz:2008sc,Lovell:2011rd,deVega:2011gg,Walker:2011zu,Laporte:2013fwa,Penarrubia:2012bb,Walker:2012td,deVega:2013ysa,Lovell:2013ola}. }   It is also possible that the sterile neutrinos make up only a fraction of dark matter~\cite{Palazzo:2007gz,Boyarsky:2008xj,Boyarsky:2008mt}, in which case they can still be responsible for the observed velocities of pulsars~\cite{Kusenko:1997sp,Kusenko:2006rh}.

\item A modification of the DW scenario proposed by Shi and
Fuller~\cite{Shi:1998km} uses a non-zero lepton asymmetry $L$.   The oscillations on 
Mikheev--Smirnov--Wolfenstein (MSW)  resonance~\cite{Wolfenstein:1977ue,Mikheev:1986gs} generate a greater abundance of relic sterile neutrinos with a lower average velocity than in the DW case.  This
results in a colder dark matter with smaller mixing angles, which relaxes the
bounds from small-scale structure and from the X-ray observations.  The
Shi--Fuller (SF) scenario works for a pre-existing lepton asymmetry $L \gtrsim10^{-3}$.  An economical model that can generate the requisite lepton asymmetry was proposed by Laine and Shaposhnikov~\cite{Laine:2008pg}: decays of the heavier sterile neutrinos could
be responsible for generating the lepton asymmetry of the universe that creates the conditions for producing dark matter in the form of the lighter sterile neutrinos. 

\item The bulk of sterile neutrinos could be produced from decays of gauge-singlet Higgs 
bosons at temperatures above the $S$ boson mass, $T\sim
100$~GeV\cite{Kusenko:2006rh}.  In this case,  the Lyman-$\alpha$ bounds on the
sterile neutrino mass are considerably weaker than in DW case or SF case because the
momenta of the sterile neutrinos are red-shifted as the universe cools down
from $T\sim 100$~GeV~\cite{Kusenko:2006rh,Petraki:2007gq,Merle:2013wta}.

\item Sterile neutrinos could be produced from their coupling to the inflaton~\cite{Shaposhnikov:2006xi} or the radion~\cite{Kadota:2007mv}.  Depending on the time of production, the population of dark-matter particles can be warm or cold.  For example if the mass of the inflaton is below 1~GeV~\cite{Shaposhnikov:2006xi}, one does not expect a significant redshifting of dark matter, which remains warm in this case.   However, if the sterile neutrinos are produced at a higher scale, they can be red-shifted as in the case of the electroweak-scale production~\cite{Kusenko:2006rh,Petraki:2007gq}.  

\item Split seesaw model~\cite{Kusenko:2010ik} allows for two different production mechanisms, both of which operate at a high energy scale. 
Both mechanisms generate a population of dark-matter sterile neutrinos that is at least as cold as it is in the case of the Higgs decays.
One of these possibilities is accompanied by an additional cooling due to entropy production in a first-order phase transition breaking $(B-L)$ gauge symmetry. 

\end{itemize}

We note that only in the first case, namely, the DW scenario,  the
dark matter abundance is directly related to the mixing angle. 
Nevertheless, the production by oscillations cannot be turned off, and the X-ray
bounds, which depend on the mixing angle, apply even in the case when only a
fraction of dark matter comes from neutrino
oscillations~\cite{Kusenko:2006rh,Palazzo:2007gz,Boyarsky:2008xj,Boyarsky:2008mt}. 
For the same reason, it is a generic prediction that dark matter should contain two components with different velocity distributions: the 
warm DW component produced at low energy, and a colder population of sterile neutrinos produced at a higher temperature and redshifted 
due to entropy production~\cite{Kusenko:2006rh,Petraki:2007gq,Boyanovsky:2007ba,Boyanovsky:2008nc,Kusenko:2010ik}.

\begin{figure}[!htb]
  \includegraphics[width=\textwidth]{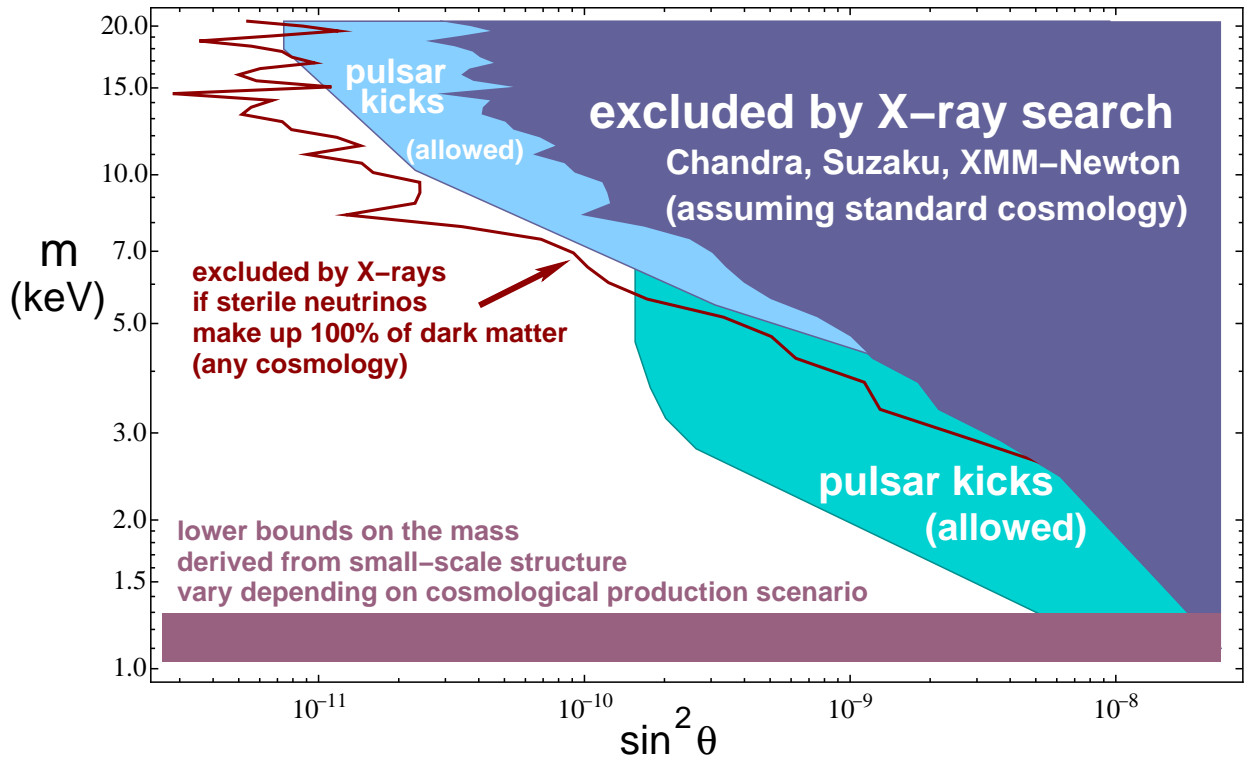}
\caption{Sterile neutrino parameters to the right of the
  solid red curve are excluded by the X-ray observations, if the sterile neutrinos 
make up all of dark matter.  If the sterile neutrino abundance is determined by neutrino 
oscillations and no other mechanism contributes, then the excluded region is smaller (shaded area).
Lower bounds from structure formation depend on the production mechanism, because they constrain the 
primordial velocity distribution whose connection to mass and mixing is model dependent. Also shown is the range 
in which the pulsar velocities can be explain by anisotropic emission of sterile neutrinos from a supernova.
\label{fig:sterile}}
\end{figure}

Sterile neutrinos with the mass and mixing angle suitable for dark matter can also be produced in a supernova explosion.  These neutrinos 
are emitted from a cooling neutron star with anisotropy determined by the magnetic field, and the resulting recoil momentum can explain the 
observed velocities of pulsars (which are magnetized rotating neutron stars)~\cite{Kusenko:1997sp,Kusenko:1998bk,Fuller:2003gy}.  
It is intriguing that the same particle can explain both dark matter and the origin or pulsar kicks~\cite{Kusenko:2009up}.  The overlapping parameter space is shown in Fig.~\ref{fig:sterile}.

Since the expected mixing angle is very small, purely laboratory experiments cannot access the relevant parameter space in the near future, although some interesting experimental approaches have been proposed~\cite{Finocchiaro:1992hy,Bezrukov:2006cy,Ando:2010ye,deVega:2011xh}. 
However, one can use X-ray telescopes to search for a line from radiative  decays 
that can occur via a one-loop diagram.  This technique, pioneered by Abazajian, Fuller, and Tucker~\cite{Abazajian:2001vt}, 
has been employed in a wide range of studies that use either archival data, or dedicated observations of dark-matter rich systems~\cite{Abazajian:2001vt,Abazajian:2005gj,Beacom:2005qv,Mapelli:2005hq,Boyarsky:2005us,Abazajian:2006yn,Boyarsky:2006fg,Boyarsky:2006jm,Watson:2006qb,Abazajian:2006jc,Boyarsky:2006ag,Boyarsky:2006hr,RiemerSorensen:2006fh,RiemerSorensen:2006pi,Boyarsky:2006kc,Boyarsky:2006jm,Yuksel:2007xh,Boyarsky:2007ay,Loewenstein:2008yi,RiemerSorensen:2009jp,Loewenstein:2009cm,Watson:2011dw,Loewenstein:2012px}. All three existing X-ray telescopes, namely Chandra, Suzaku, and XMM-Newton have been employed for this purpose~\cite{Loewenstein:2008yi,Loewenstein:2009cm,Loewenstein:2012px}. Future X-ray instruments, such as Astro-H, provide an opportunity to explore the entire best-motivated range of masses and mixing angles, including the range in which the same particle explains both dark matter and the pulsar velocities.

In summary, sterile neutrinos present a well-motivated dark matter candidate for which indirect detection (using X-ray telescopes) offers the most promising avenue for discovery.

\subsection{Superheavy dark matter} 

In addition to primordial black holes, there are a number of dark matter candidates that have large masses and, therefore, are expected to 
have very low number densities.  The search strategies for these dark matter candidates are different from the usual searches
in that no laboratory experiment has big enough acceptance to detect a sufficient number of events,
even if these particles are strongly interacting.  Detection is nevertheless possible with the use of 
ingenious alternative techniques: for example, one can study tracks in mica (which has small size but $\sim$billion years of exposure), or seismic detectors, or ultrahigh-energy cosmic rays from massive particle decays.  Direct detection of supermassive particles is possible with the use of large-volume detectors, such as ANITA, HAWC, IceCube, Pierre Auger, Super-Kamiokande.

\subsubsection{WIMPzillas}

Extremely heavy particles can be produced in the early universe due to gravitational interactions, even if their other interactions are very weak. These particles can be created at the end of inflation~\cite{Chung:1998ua,Chung:1998zb}, and they can have $\sim 10^{13}$~GeV or even transplanckian masses~\cite{Kolb:2007vd}.  

\subsubsection{Strangelets, quark nuggets}

Extremely dense nuggets of quark matter could form im the course of a first-order phase transition and could be stable

\begin{figure}[!htb]
\begin{center}
\includegraphics[width=9cm]{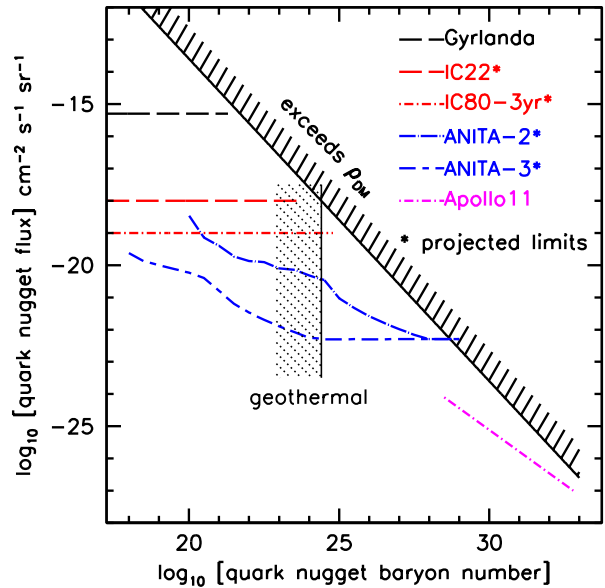}
\caption{Current and projected limits on quark nuggets.  See Refs.~\cite{Gorham:2012hy,Lawson:2013bya} for discussion.
\label{fig:QNlimits}}
\end{center}
\end{figure}

It was pointed out by Witten~\cite{Witten:1984rs} that a first-order QCD phase transition could result in formation of extremely dense stable objects 
composed of quarks.  It has also been suggested that nuggets of both matter and antimatter are formed as a result of the dynamics of the axion domain walls \cite{Zhitnitsky:2002qa,Oaknin:2003uv}. The dynamics of their formation is not well understood, and the absence of a first-order phase transition in QCD at high temperature established in recent studies~\cite{Aoki:2006we} calls into question some of the scenarios considered earlier.  If stable quark and antiquark nuggets can form in the early universe, some large-effective-volume detectors, such as ANITA~\cite{Gorham:2012hy} can provide experimental bounds, as shown in Fig.~\ref{fig:QNlimits}. 

\subsubsection{Indirect detection of superheavy dark matter}

One can detect cosmic rays, photons, and neutrinos form decays of superheavy dark matter particles. Their annihilation is not important because the large masses imply low number densities.  A spectacular signature could be ultrahigh-energy cosmic rays coming from decays of dark-matter particles.  Due to inevitable QCD fragmentation, decays of superheavy particles should be accompanied by a flux of ultrahigh-energy photons~\cite{Gelmini:2005wu}.  The upper limits on such photons set by the Pierre Auger Observatory~\cite{Abraham:2009qb} has ruled out the possibility that all UHECR above 30 EeV could come from superheavy dark matter decays.  However, WIMPzillas can make up a small fraction of dark matter, or they can still account for all dark matter if they are stable.
PeV neutrinos recently 
discovered by IceCube~\cite{Aartsen:2013bka} have been linked to the possibility of superheavy dark matter~\cite{Feldstein:2013kka,Esmaili:2013gha}.

In summary, superheavy dark matter can be discovered using large-size detectors.

\subsection{Supersymmetric Q-balls and the products of their decays} 
Supersymmetry (SUSY) is an elegant and plausible extension of the Standard Model. Thanks to 
the presence of scalar fields endowed with conserved baryon and lepton numbers, the existence of stable or unstable $Q$-balls~\cite{Coleman:1985ki} is generic in any supersymmetric generalization of the Standard Model~\cite{Kusenko:1997zq}. Non-topological solitons exist in any theory with (i) scalar fields carrying a global charge $Q$ associated with some global $U(1)$ symmetry, and (2) a scalar potential $V(\phi)$ such that  
\begin{equation}
\frac{V(\phi)}{\phi^{2}} = {\rm min}, ~~{\rm for}~~\phi = \phi_{0} \neq0, 
\label{eqQball}
\end{equation}
where $\phi_0$ is the minimum of the potential.

The first condition is met by the presence of squarks and sleptons, which carry baryon or lepton number. The second condition (\ref{eqQball}) is satisfied by both the presence of trilinear terms and by the so-called ``flat directions'' in the scalar potential. The flat are directions in scalar field space for which, in the limit of exact supersymmetry, the potential vanishes identically at the renormalizable level.\footnote{These flat directions are well studied and catalogued in the case of the minimal supersymmetric standard model (MSSM)~\cite{Gherghetta:1995dv}.} Hence the only contributions to the scalar potential along these directions come from higher-dimensional operators and terms originating from SUSY-breaking. Thus one expects the $Q$-ball solution to be quite sensitive to the nature of SUSY-breaking, which is indeed the case: gauge-mediated SUSY-breaking (where the potential is very flat) tend to predict stable $Q$-balls, whereas models of gravity-mediation tend to predict unstable $Q$-balls.  
These stable $Q$-balls appearing in gauged-mediated models offer an attractive dark matter candidate~\cite{Kusenko:1997si}.  However, even when $Q$-balls are unstable, they are typically so long-lived that they may source the baryon asymmetry and/or offer a non-thermal production mechanism of stable dark matter LSPs~\cite{Enqvist:1997si,Enqvist:1998en,Roszkowski:2006kw,Shoemaker:2009kg}.

\subsubsection{Production of SUSY $Q$-balls from Affleck-Dine condensate fragmentation}

A combination of supersymmetry and inflation creates a natural mechanism for the abundant creation of very large $Q$-balls through the Aflleck-Dine mechanism of baryogenesis~\cite{Affleck:1984fy,Dine:1995uk,Dine:1995kz} (see Ref.~\cite{Dine:2003ax} for a review).

Here we will briefly sketch the evolution of the flat direction field $\phi$ in the early Universe. Let us first recall that for a coherent scalar field to acquire a large charge density, $n_{\phi}\sim \dot{\theta} |\phi|^{2}$, one requires (1) a large field value and (2) nonzero angular motion.  

During inflation, the large energy density from the inflaton field breaks supersymmetry, and generates a mass-squared term for the flat direction field, $\pm H^{2}|\phi|^{2}$, where $H$ is the Hubble parameter. The sign of this term is model-dependent, but when it is negative, the flat direction field $\phi$ naturally develops a very large value that is only stabilized by higher-dimensional operators, $W_{n} = \phi^{n+3}/\Lambda^{n}$, at a field value 
\begin{equation}
\phi_{0} \approx c_{n}~ H^{\frac{1}{1+n}} \Lambda^{\frac{n}{n+1}}
\end{equation}
where $c_{n}$ is a $\mathcal{O}(1)$. In slow-roll inflation, the field $\phi$ can continuously find the new minimum of the potential even after inflation ends and the Hubble parameter begins to decline. 

A nonzero angular motion for $\phi$ is generated by the two $B$-violating terms, $V_{\not{B}} \ni a m_{3/2} W_{n} + b H W_{n}$. The sign of the relative $CP$-violating phase between the coefficients of these terms sets the handedness of the rotation, i.e. determining whether a net baryon or anti-baryon number is produced. Having produced a net angular rotation with a large VEV, this mechanism generates a charge asymmetry inside the scalar condensate. A detailed calculation of the asymmetry can be obtained by examining the equations of motion for $\phi$ and noting that the epoch of asymmetry generation occurs when the two  $B$-violating terms are comparable. 

In addition to the bulk evolution of the condensate, one can also perform a stability analysis, and track the evolution of perturbations on top of the bulk motion. In Ref.~\cite{Kusenko:1997si},  it was shown that, when the condensate carries a sufficiently large charge, an instability sets in to destroy the homogeneity of the condensate. This instability breaks up the initially homogeneous condensate into lumps of charge which rapidly evolve to find their ground states: $Q$-balls.\footnote{We note however, that the true ground state configuration may occur off the flat direction, 
resulting in an emission of baryon/lepton number and a mechanism for acquiring electric charge~\cite{Shoemaker:2008gs}.} In some instances, this fragmentation may be so violent as to produce a potentially detectable gravitational wave signal~\cite{Kusenko:2008zm,Kusenko:2009cv,Chiba:2009zu,Allahverdi:2012ju}.

\subsubsection{Properties of SUSY $Q$-balls and their Detection}
In models of gauge-mediated supersymmetry breaking, the scalar potential grows quadratically at field values small compared to the SUSY-breaking scale $M_{S}$, and then flattens to mere logarithmic growth above it. In such a potential, the $Q$-ball solution can be shown to have a rest mass scaling as  $M(Q_{B}) \sim M_{S} Q_{B}^{3/4}$, where $Q_{B}$ is the baryonic charge of the $Q$-ball. The fact that the $Q$-ball mass grows less quickly than 
$Q_B$ is crucial to its stability. We recall that the $Q$-ball solution is the scalar field configuration that minimizes the energy for a fixed amount of charge $Q_{B}$. Thus the $Q$-ball is, by construction, stable with respect to decay into scalars but not necessarily fermions. Decay into nucleons for a baryonic $Q$-ball is impossible if 
\begin{equation}
\frac{M(Q_{B})}{Q_{B}} \sim \frac{M_{S}}{Q_{B}^{1/4}} < m_{p}.
\end{equation}
Therefore, $Q$-balls with baryonic charge $Q_{B} > 10^{12}\left(M_{S}/{\rm TeV}\right)^{4}$ are stable dark matter candidates.  These stable $Q$-balls carry a large baryon number and have large cross sections, given by their geometric size:  
\begin{equation}
\sigma_{Q} = \frac{\pi}{2}\frac{Q_{B}^{1/2}}{M_{S}^{2}} \approx 600~{\rm barn} \left(\frac{Q_{B}}{10^{24}}\right)^{1/2}\left(\frac{1~{\rm TeV}}{M_{S}}\right)^{2}.
\end{equation}
Despite such large cross sections, SUSY $Q$-balls remain challenging experimental targets~\cite{Takenaga:2006nr} because of their low number density. 

For baryonic number $Q_B\sim 10^{26}$, which is typical for Affleck--Dine baryogenesis, one obtains the correct dark matter abundance and the correct baryon asymmetry of the universe~\cite{Laine:1998rg}.

Let us now consider the details of $Q$-ball interactions with matter. When a $Q$-ball enters the Earth's atmosphere and encounters its first nucleon, the quarks of the nucleon scatter off the large squark VEV inside the $Q$-ball. In SUSY there is a coupling $\tilde{q}q\tilde{g}$ between gluinos, squarks and quarks. Thus inside the $Q$-ball the quarks get a large Majorana mass as they mix with the gluinos, $\langle \tilde{q}\rangle q\tilde{g}$, and this Majorana mass term violates the baryon number conservation. A quark scattering off a Q-ball has a large probability to reflect as an antiquark~\cite{Kusenko:2004yw}.  This allows the roughly 1 GeV of QCD binding energy inside the nucleon to be converted into a burst of pions. This pion signal forms the basis of the main experimental detection strategy for electrically neutral SUSY $Q$-balls~\cite{Takenaga:2006nr}. 

In contrast with WIMP dark matter, $Q$-balls do not accumulate inside the Sun or Earth. This is because, despite their large interaction strength with ordinary matter, the momentum is so large that numerous scatterings are insufficient to slow their motion below the escape speed of stars and planets. The interactions of $Q$-balls can instead be probed in three complementary ways: (i) $Q$-balls can be stopped by dense objects like white dwarfs and neutron stars, with potentially dramatic consequences for the evolution of such stars~\cite{Kusenko:1997it,Kusenko:2005du}; (ii) the passage of $Q$-balls through a large-area detector, such as Super-Kamiokande or HAWC may be detected~\cite{Kusenko:1997vp,Kusenko:2004yw}; (iii)  the spectrum of neutrinos produced by $Q$-balls passing through the entire Earth may be detectable~\cite{Kusenko:2009iz}.

\subsubsection{Dark Matter from $Q$-ball Decays}
In Refs.~\cite{Enqvist:1997si,Enqvist:1998en} it was pointed out that $Q$-balls offer a  natural mechanism for the simultaneous production of dark matter and the baryonic ordinary matter. This is due to the fact that the decay of squarks inside the $Q$-ball via $\tilde{q} \rightarrow q + {\rm LSP}$ can occur much later than a similar decay in the plasma. The decay of $Q$-balls into fermions becomes Pauli-blocked as the fermions populate the interior of the $Q$-ball~\cite{Cohen:1986ct}. For every unit of baryon number lost at least three LSPs are generated, $N_{LSP} \ge 3$. Thus for $Q$-balls which has trapped all of the produced baryon number one finds that the amounts of dark matter and baryons are related via
\begin{equation}
\frac{\Omega_{\rm B}}{\Omega_{\rm DM}} \sim \frac{m_{n}}{N_{\rm LSP}\ m_{\rm LSP}},
\end{equation}
where $m_{n}$ is the mass of the neutron. Clearly, this mechanism prefers GeV-scale LSP dark matter. The original gravity-mediated scenario~\cite{Enqvist:1997si,Enqvist:1998en} with GeV mass neutralino dark matter is now excluded by LEP searches. 
However gauge mediation is a natural to consider in this context because it already has a light LSP, the gravitino. For example, for low reheat temperatures, $Q$-balls with charges in the range $10^{12} \le Q \le 10^{18}$ decay after reheating and can simultaneously account for the observed dark-matter--to--baryon ratio~\cite{Shoemaker:2009kg}.  
Recent work on gravitino dark matter from $Q$-ball decays have highlighted the the need for a large messenger masses~\cite{Doddato:2011fz}, the sensitivity of BBN constraints to the nature of the NLSP~\cite{Kasuya:2011ix,Doddato:2011gd}, and the  importance of NLSP decays/annihilations in determining the final gravitino abundance~\cite{Kawasaki:2012gk,Kasuya:2012mh}. Furthermore, gravity mediated supersymmetry offers new interesting possibilities for pangenesis~\cite{Kamada:2012bk}.

\subsection{Supersymmetry's non-WIMP candidates and other non-WIMP candidates}

Supersymmetric extensions of the Standard Model provide a well-motivated and appealing framework for new physics beyond. In addition to 
SUSY WIMPs, and SUSY Q-balls discussed above, supersymmetry allows other possibilities for dark matter. 

\subsubsection{Supersymmetry and axion}

In theories with supersymmetry  the
Peccei-Quinn  solution to the strong CP problem brings into existence both the axion and its supersymmetric partner.  
In this case, the PQ axion field is promoted to a chiral superfield which 
contains as well an $R$-parity even spin-0 saxion $s$ and an $R$-odd
spin-$1/2$ axino $tilde{a}$. Supergravity mass calculations typically lead 
to the saxion and the axino both with masses of order the gravitino mass  
$m_{3/2}$, which is
taken to be around the TeV-scale in gravity-mediated SUSY breaking models.
In such a case, then one can expect {\it two} dark matter particles: the axion
along with the lightest neutralino.  
In mixed axion/neutralino CDM models, it may be possible to detect both a WIMP and an axion as dark matter relics~\cite{Baer:2009vr,Baer:2011hx}.

\begin{figure}[!htb]
\begin{center}
\includegraphics[width=0.7 \textwidth]{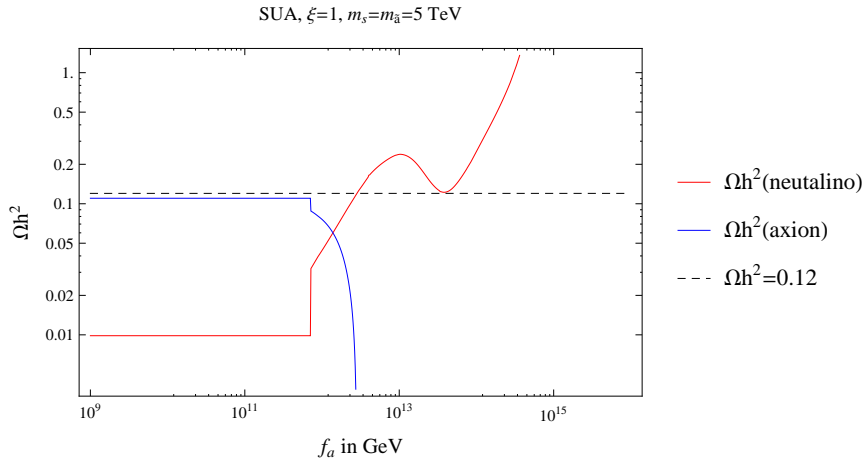}
\caption{Neutralino and axion relic abundance versus $f_a$ 
in the SUSY DFSZ model. See Refs.~\cite{Baer:2009vr,Baer:2011hx} for discussion.
\label{fig:SUAxi=1}}
\end{center}
\end{figure}

\subsubsection{SuperWIMP Models}

Both WIMPs and superWIMPs emerge naturally in several well-motivated particle physics frameworks, such as
supersymmetry, universal extra dimensions and brane-worlds.
However, the theoretical implications of superWIMPs are completely different from those of WIMPs.
We can illustrate this fact with $R$-parity conserving supersymmetry models, in which the lightest
supersymmetric particle (LSP) is completely stable. Within the WIMP scenario, the slepton LSP region
of the parameter space is excluded cosmologically.  In most part of the remaining allowed region,
the neutralino is the LSP.  Much of the neutralino LSP region is excluded because neutralinos are
overproduced. The situation is different within the superWIMP framework, where, for example,
the axino or the gravitino is the lightest supersymmetric particle (LSP). The regions of parameter space
where a slepton is the lightest SM superpartner are especially interesting, since late decays
to gravitinos can impact Big Bang nucleosynthesis (BBN) and possibly even resolve the anomalies associated with the primordial abundance of 
$^7$Li~\cite{Feng:2003xh,Kawasaki:2004qu,Pospelov:2006sc,Kaplinghat:2006qr}. Following the same argument, much of the region with neutralino as the next lightest supersymmetric partner (NLSP) is disfavored for the gravitino LSP case, since the hadronic neutralino decay typically destroys BBN successes.
On the other hand, regions excluded by overproduction within the classical WIMP framework, are the most interesting within the superWIMP scenario. In the standard case, where a WIMP decay produces one superWIMP, the abundance of the dark matter is reduced by the ratio of WIMP to superWIMP masses:

\begin{equation}
\Omega_\text{SWIMP}=\frac{m_\text{SWIMP}}{m_\text{WIMP}}\,\Omega_\text{WIMP}\,.
\label{omegas}
\end{equation}

Another possibility within this scenario is that of mixed warm and cold DM models.
A particularly interesting example arises when the axino is the LSP. The phenomenology
of such a superWIMP is characterized by a short WIMP lifetime compared to the gravitino LSP case. These models have received some attention in the last few years since they are less constrained by astrophysical observations~\cite{Seto:2007ym}. Indeed, DM composed of a mixture of axion and axino has been claimed to be favored in simple supersymmetric constructions \cite{Baer:2009vr}.

\subsubsection{SuperWIMP signatures}

SuperWIMPs signatures have attracted a lot of interest from different points of view~\cite{Covi:1999ty,Feng:2003nr,Feng:2003xh}.
This idea has provided new search strategies at colliders~\cite{Feng:2005gj,Hamaguchi:2004df,Feng:2004yi,Cembranos:2006gt}.
Depending on the charge of the decaying particle and its lifetime, superWIMP scenarios provide a rich
variety of exotic collider signals, such as displaced vertices, track kinks, tracks with nonvanishing impact parameters,
slow charged particles, and vanishing charged tracks \cite{Cembranos:2006gt}.
The decay lifetime could be months or even years. Given this, there have been different proposals to trap these particles (if charged) outside of the particle detector so that their decays can be analyzed and characterized~\cite{Feng:2004yi,Hamaguchi:2004df}.

On the other hand, the superWIMPs can leave their imprint on early universe cosmology. For example, the primordial element abundances may be modified due to energy injections from late time decays~\cite{Kawasaki:2004qu} and due to formation of new bound states with new meta-stable charged particles~\cite{Pospelov:2006sc,Kaplinghat:2006qr}. These analyses can be used to constrain superWIMP models, but as we have commented above, in some regions of parameter space they may explain present inconsistencies in the Lithium abundance. This possibility can be corroborated. For instance, the heavy meta-stable charged particles could be produced in cosmic rays and detected with high energy neutrino telescopes or in sea water experiments \cite{Albuquerque:2003mi,Bi:2004ys}. In addition, late decays could also distort the Blackbody spectrum of the cosmic microwave background \cite{Yeung:2012ya} or be detected directly by studying cosmic ray spectra \cite{Cembranos:2007fj,Garny:2010eg}.

\begin{figure}[!htb]
  \includegraphics[width=0.44\textwidth, clip=False, trim = 1mm 1mm 1mm 1mm]{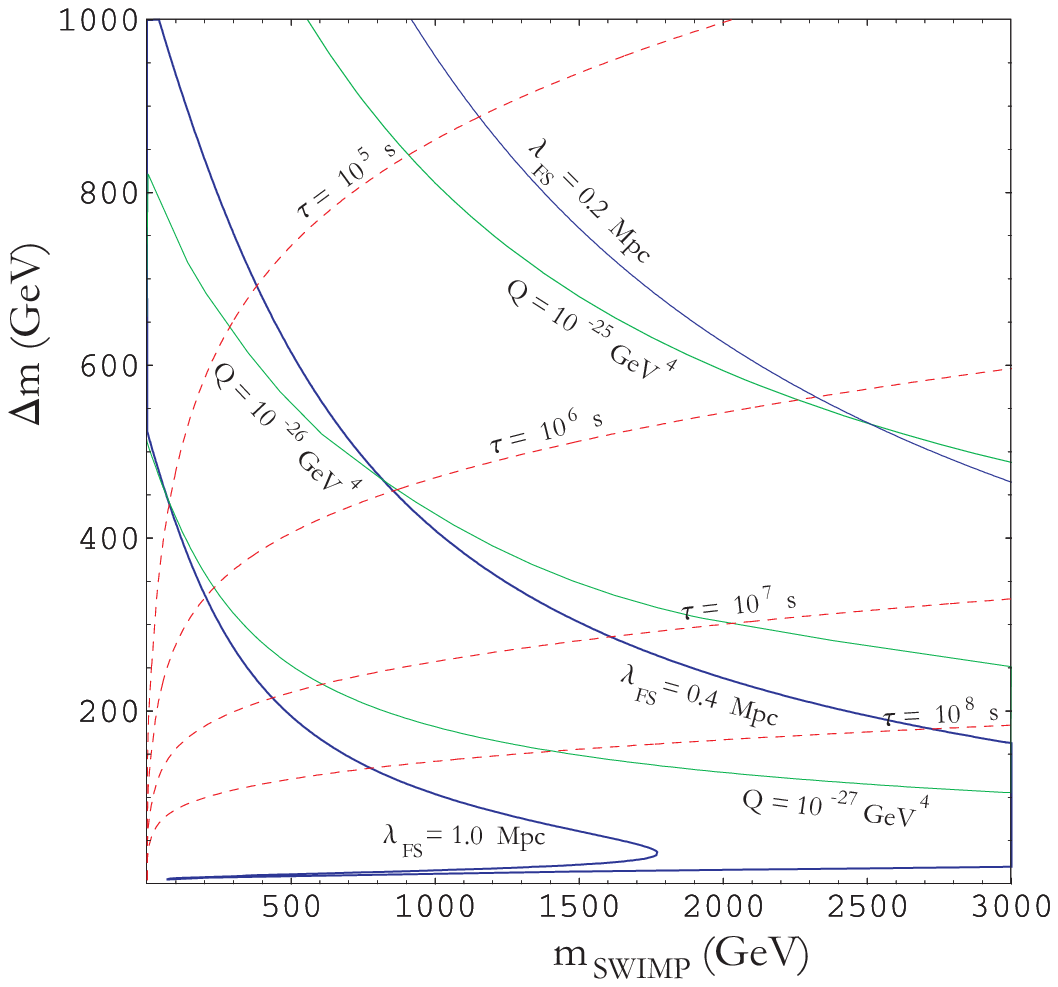}
  \includegraphics[width=0.45\textwidth, clip=True, trim = 10mm 10mm 10mm 25mm]{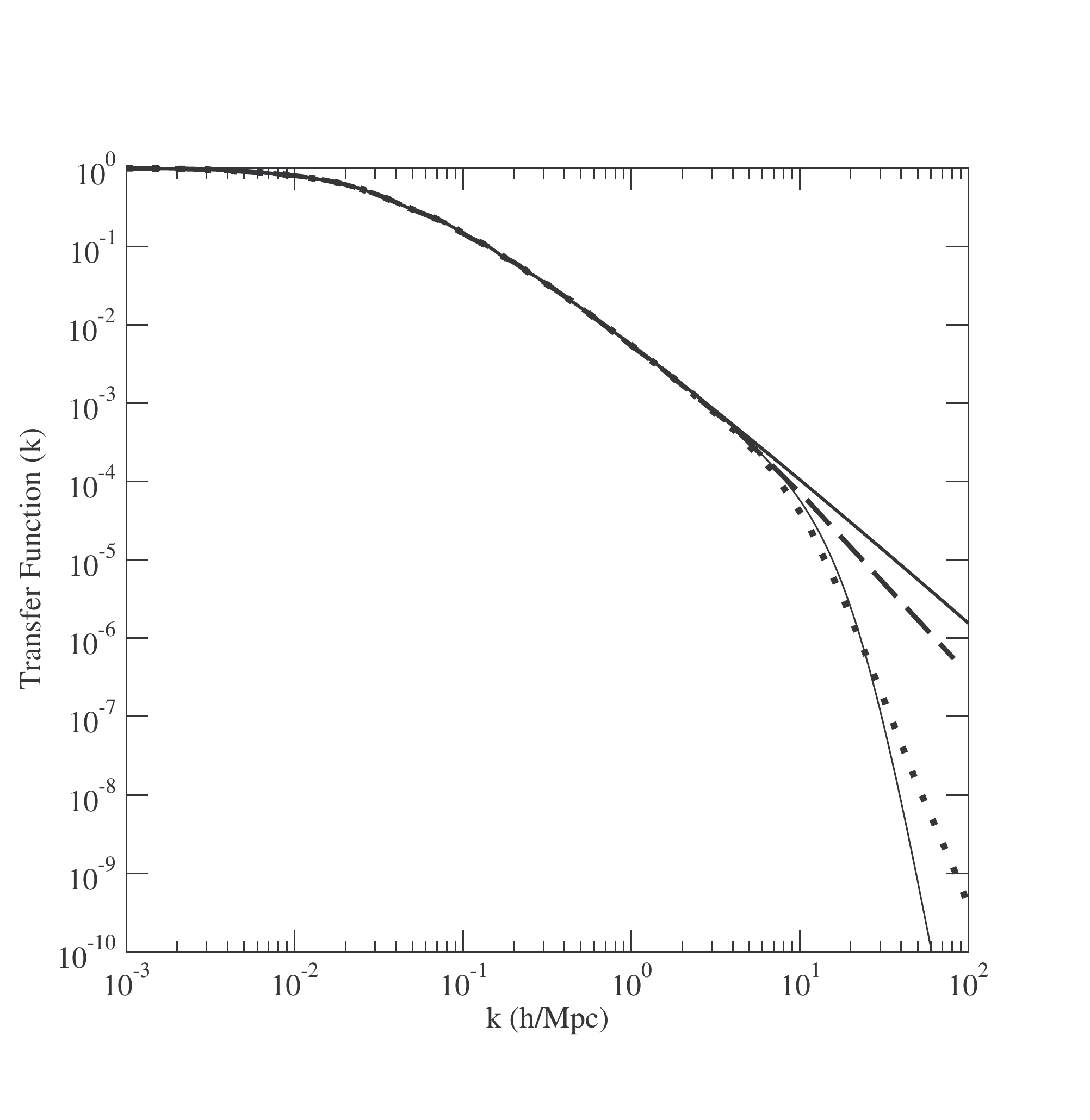}
\caption{{\bf Left}: SuperWIMP parameter space  $(m_\text{SWIMP}, \Delta m)$, where
$\Delta m \equiv m_\text{WIMP} - m_\text{SWIMP}$ for gravitino LSP (superWIMP)
and sneutrino NLSP (WIMP). $Q$ is the phase space density as defined by \cite{Hogan:2000bv}, $\lambda_\text{FS}$ is the free-streaming scale, and $\tau$ is the lifetime of the sneutrino. On the bottom-left corner, superWIMP DM behaves as
hot DM and it is excluded. On the top-right corner, it behaves as cold DM. Between both regimes, superWIMPs
work as a new type of warm DM that could reduce central densities and substructure in observable ways.
{\bf Right}: The curves show the power spectra for different values of $f$, the fraction of dark matter today that arises from decays as opposed to those produced during reheating (which would be cold dark matter). The solid curve shows the $f=0$ case (CDM). The dashed curve shows the $f=0.5$ case while the dotted curve shows the $f=1$ case. It is clear that the suppression on small scales is much reduced for the $f=0.5$ case. For comparison, we also plot (see thin solid curve) the transfer function for a 1 keV thermal Warm Dark Matter model.}
\label{sph}
\end{figure}

Despite their large masses, superWIMPs could behave (effectively) as warm dark matter (WDM) \cite{Cembranos:2005us,Kaplinghat:2005sy}. In fact, depending on the lifetime and the kinetic energy associated to the decay, they can work as hot, warm, cold or meta-cold DM. There are various puzzles in galaxy formation and one of the puzzles that has garnered much attention is the issue of the missing satellites \cite{Klypin:1999uc,Moore:1999nt}, essentially the question of how the thousands of subhalos in CDM simulations can be reconciled with the small number of Milky Way satellites discovered (about 20). In WDM models the formation of small mass halos is suppressed and hence it has been proposed as a solution for the missing satellites problem. Recent work has pointed out a new issue with Milky Way satellites -- the observed bright satellites are underdense in dark matter compared to the most massive subhalos of a Milky Way halo in CDM simulations \cite{BoylanKolchin:2011dk}. This is puzzling because it is 
expected that the most massive subhalos would host the bright satellites. Recent work has claimed that this issue is also solved in WDM models because subhalos have lower densities compared to their CDM counterparts \cite{Lovell:2011rd} due to the lack of power on small scales. The required WDM solution can be found within the context of superWIMP models by estimating the power spectrum cut-off scale~\cite{Sigurdson:2003vy,Cembranos:2005us,Kaplinghat:2005sy,Strigari:2006jf}. This is shown in more detail in Fig. \ref{sph} and we estimate that the region between the free-streaming scales of 0.2-0.4 Mpc could be the relevant WDM solution.

To summarize, superWIMPs arise in well-motivated theoretical frameworks of beyond standard model physics.
These particles can inherit the relic abundance of WIMPs since they arise from the decay of a WIMP. The interactions of superWIMPs with standard model particles are strongly suppressed. However, there is a rich variety of distinctive signatures in this scenario at colliders and in astrophysics and cosmology.

\subsubsection{Supersymmetry/string moduli}

Light scalar fields called moduli arise from a variety of different models involving supersymmetry
and/or string theory; thus, their existence is a generic prediction of leading theories for physics
beyond the standard model. These fields also present a formidable, long-standing problem for cosmology.
An anthropic solution to the moduli problem leads to dark matter in the form of moduli particles. 
This form of dark matter is consistent with the observed properties of structure formation, and it is amenable
to detection with the help of X-ray telescopes~\cite{Kusenko:2012ch}.

\section{Concluding remarks}

The family of non-WIMP dark-matter candidates is very large. Some candidates have been under intense study,
both theoretically and experimentally, and some are
explored with lesser enthusiasm. Some candidates are perceived as more motivated than the other.  For example, the axion 
has the advantage of being a natural solution to a strong CP problem.  Furthermore, 
for a dark-matter candidate, the QCD axion is special in that it has a fairly well bounded parameter
space. The axion-photon couplings $g_{a\gamma\gamma}$ over the range of benchmark models
extend over an order of magnitude. The upper end of the QCD axion mass range is set at a few milli-eV by the
limit from SN1987A (though there might be ways to evade this), the lower end, limited by the ``misalignment''
cosmological bound, is set at around a $\mu$eV (though, as discussed about, there are ways to evade this bound).
The most promising approach to detecting the QCD axion is with the RF-cavity technique.
Although the expected conversion into RF power within the cavity is extraordinarily weak, experiments will shortly
start taking data for a definitive search. By ``definitive'', we mean one that will either find the axion with high confidence, if it exists,
or if not, exclude it at high confidence. These experiments will sensitively explore the first two decades of allowed QCD axion mass where
the dark-matter QCD axion is expected to be. Starkly, these searches have large discovery potential.
There are axion and axion-like-particle alternatives to the QCD axion, and this
opens a vast and largely unexplored search space. Much of this
space, including the third decade of allowed mass for the QCD axion, would be sensitively explored by IAXO,
should IAXO be supported. The large region open to IAXO for non-QCD axions should as well give these searches
a good discovery potential.
The axion might be found anywhere within the open parameter space. There are
proponents of both higher-mass axions (which would then be of the non-PQ type) or lower-mass axions (which
could be QCD axions of the ``anthropic'' type, for instance.)
It may also be that the dark matter consists of one of the other dark matter candidates, WIMP or non-WIMP, or
a mixture of candidates. There are an enormous number of possibilities. 

Aside from axions, special-purpose proposed searches for non-WIMP
dark matter are less well developed.  They are nevertheless conducted, using serendipitous capabilities of existing experiments. 
For example, X-ray telescopes are used to search for relic sterile neutrinos, Super-Kamiokande has produced limits on SUSY Q-balls, and 
gamma-ray telescopes are used to constrain some forms of asymmetric dark matter.  However, a careful study of possibilities for 
dedicated non-WIMP dark matter detectors would be worthwhile. 

Finally, looking into the future, we believe that when the axion or other dark-matter particle is identified, it will mark a new beginning.
For instance, one virtue of the RF-cavity experiments is that they measure the total energy of the axion, mass plus kinetic,
and there may be fine structure to the signal due to the flows of dark matter in the halo; this contains a wealth of
information about the history of the formation of our Milky Way galaxy and will mark the beginning of a new field of astronomy.
If dark matter is made up of sterile neutrinos, the narrow spectral line from their decay could provide information about the 
redshift, allowing one to map out dark matter in the universe and to use the redshift information to study cosmological expansion.
Much the same can be said about discovering any of the dark matter candidates: the identification of dark matter
would be a revolutionary discovery that will open the door to a new chapter in our understanding of nature.

\bibliography{nonwimpbib}

\end{document}